\documentclass[12pt]{article}
\usepackage[margin=2 cm]{geometry}
\usepackage{comment}
\usepackage{graphicx}
\usepackage{amsmath}
\usepackage[font={small,it}]{caption}
\usepackage{mwe}
\usepackage[nottoc]{tocbibind}
\usepackage{amsmath,amssymb,extarrows,mathtools,graphicx,subfigure,setspace}
\usepackage{cite}
\usepackage{stackengine}
\usepackage{tikz}\usetikzlibrary{calc}
\usepackage{braket}
\usepackage{epsfig} 
\usepackage[section]{placeins}
\usepackage{slashed}
\usepackage{color}
\usepackage{caption}
\usepackage{amsmath}
\usepackage{hyperref}
\makeatother
\newmuskip\pFqmuskip

\newcommand{\be}{\begin{equation}}
\newcommand{\bea}{\begin{eqnarray}}
\newcommand{\eea}{\end{eqnarray}}
\newcommand{\ba}{\begin{array}}
\newcommand{\ea}{\end{array}}
\newcommand{\ee}{\end{equation}}
\newcommand{\bes}{\begin{equation*}}
\newcommand{\beas}{\begin{eqnarray*}}
\newcommand{\eeas}{\end{eqnarray*}}
\newcommand{\bas}{\begin{array*}}
\newcommand{\eas}{\end{array*}}
\newcommand{\ees}{\end{equation*}}

\newcommand*\pFq[6][8]{%
  \begingroup 
  \pFqmuskip=#1mu\relax
  \mathcode`\,=\string"8000
  \begingroup\lccode`\~=`\,
  \lowercase{\endgroup\let~}\pFqcomma
  {}_{#2}F_{#3}{\left[\genfrac..{0pt}{}{#4}{#5};#6\right]}%
  \endgroup
}
\newcommand{\pFqcomma}{\mskip\pFqmuskip}

\numberwithin{equation}{section}

\begin{document}

\onehalfspacing
\vfill
\begin{titlepage}
\vspace{10mm}

\begin{center}

\vspace*{10mm}
\vspace*{1mm}
{\Large  \textbf{Some universalities in the partition functions of low-dimensional gravity models}} 
 \vspace*{1cm}
 
 {$\text{Mahdis Ghodrati}^{a,b}$}
 
 \vspace*{8mm}

{ \textsl{ $^a $ International Centre for Theoretical Physics Asia-Pacific,
University of Chinese Academy of Sciences, 100190 Beijing, China}} \\

{ \textsl{ $^b $Asia Pacific Center for Theoretical Physics, Pohang University of Science and Technology,
Pohang 37673, Republic of Korea}}

 \vspace*{0.7cm}

\textsl{e-mails: {\href{mahdisghodrati@ucas.ac.cn}{mahdisghodrati@ucas.ac.cn}}}
 \vspace*{2mm}

\vspace*{1.7cm}

\end{center}

\begin{abstract}

In this work, first, we discuss the connections between various low-dimensional quantum gravity models, including 3d Chern-Simons, 2d Jackiw-Teitelboim (JT),  2d background field (BF) theory, 2d Liouville, 2d Wess-Zumino-Witten (WZW), and 1d Schwarzian, which are related through holography and dimension reduction, and discuss some universalities in their partition functions. Then, we specifically examine the JT partition function and the partition function of $\mathcal{N}=(2,2)$ on $S^2$ and $\text{AdS}_2$ and discuss their similarities and therefore examine our proposed universalities. We change the parameters in each model and based on the change in the structure of the partition functions, strengthen our conjectures. We also use eigenfunctions, spectra and the behaviors of Wheeler-DeWitt wavefunctions to generate more universalities between these low-dimensional quantum gravity models, specifically in their partition functions. Finally, we use entanglement entropy, complexity and RG flows, particularly in the context of wormholes, to find more universalities in quantum gravity models.

 \end{abstract}

\end{titlepage}

\tableofcontents

\section{Introduction}\label{sec:intro}

There are many universalities in the behavior of quantum gravity models and especially in their partition functions. One of the main universalities in gravity models is the universal structure of the near horizon of near-extremal black holes, which is an $\text{AdS}_2$ throat with a slowly varying internal space. This universal structure then can show its effects on the full partition functions of these models. The low energy gravitational dynamics also can be captured by the effective Einstein-Hilbert action plus the Jackiw-Teitelboim (JT) term, in which the size of the internal space can be captured by the dilaton field $\phi+\phi_0$. The dynamics of the rigid $2d$ AdS part comes from extrinsic curvature $K$ which couples to $\phi_b$.

Another reason that one can find such universalities is that in all of these cases, the matter only couples to the metric and not to the dilaton field, as the matter comes from the Kaluza-Klein reduction. Due to this fact, we can separate the partition functions into two distinct sections in many of the gravity models where these two terms have very different behaviors.

\begin{figure}[ht!]
 \centering
   \includegraphics[width=8cm] {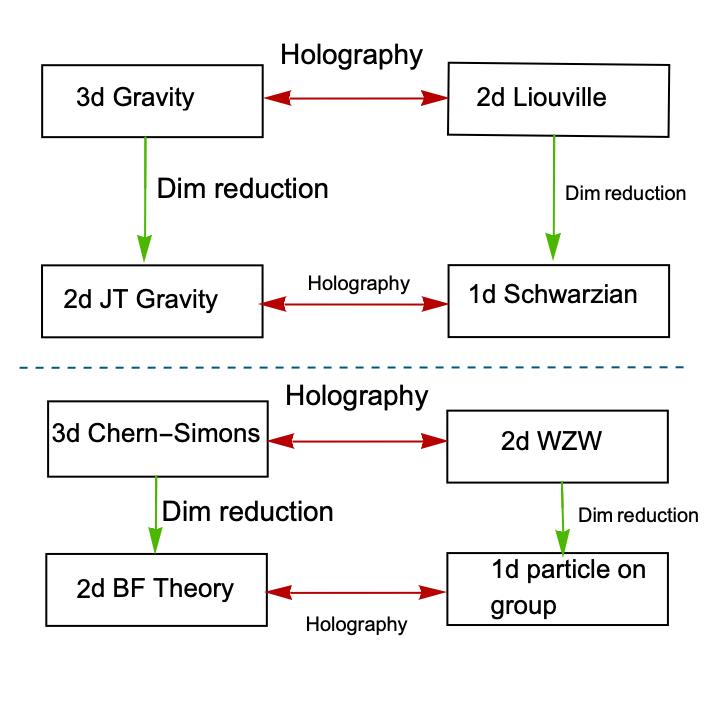}  \ \ \
  \includegraphics[width=8cm] {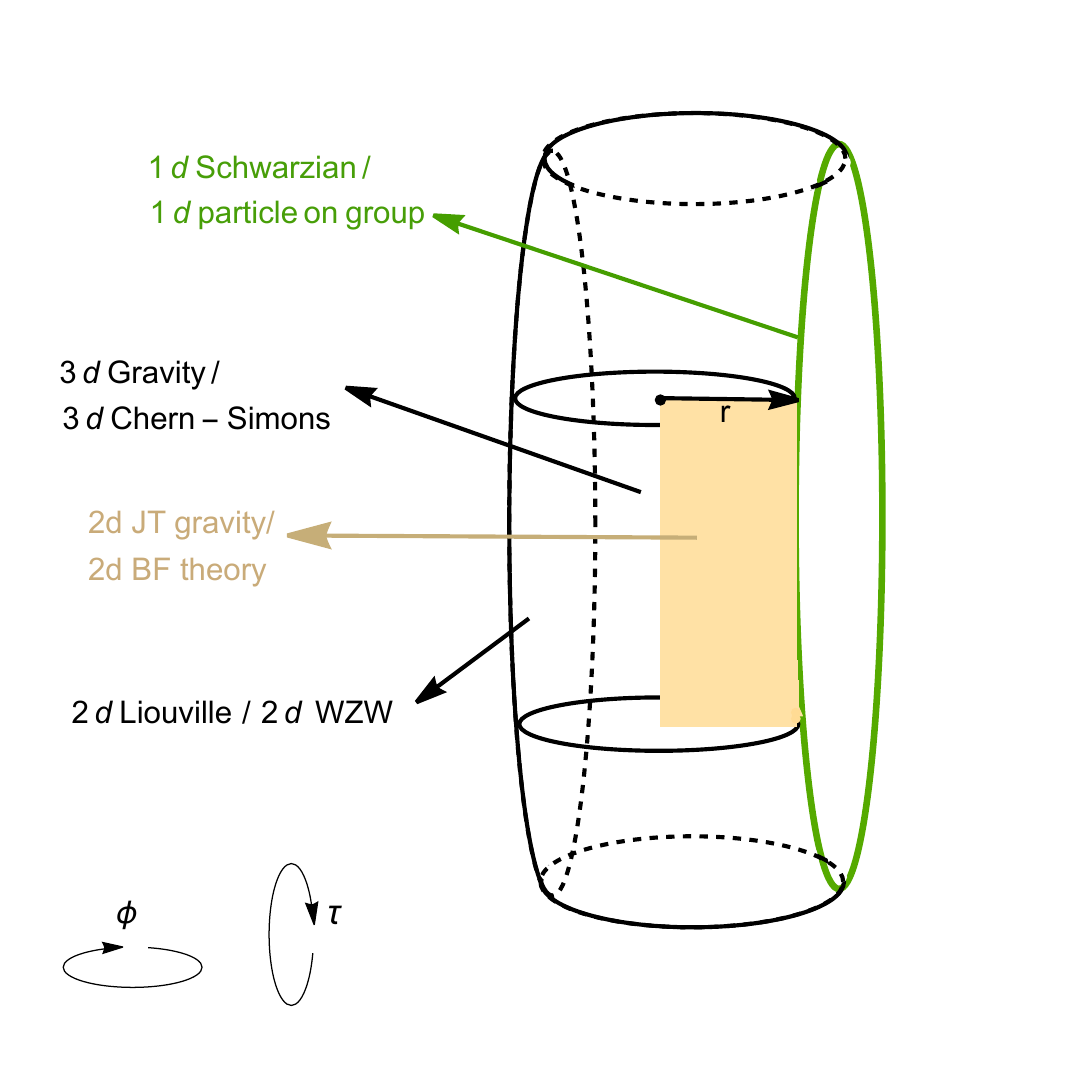} 
  \caption{The connections between various low-dimensional ($2d$ and $3d$) gravity models. This figure is based on the work of \cite{Mertens:2018fds,Blommaert:2018oro}, which shows the relations among three-dimensional gravity/Chern-Simons theory, Liouville/WZW theory, JT/BF theory and Schawarzian/particle-on-a-group description.}
 \label{fig:dimHoltheories}
\end{figure}

These low-dimensional gravity models have also interesting connections among themselves, through dimension reduction and holography, as shown in Figure \ref{fig:dimHoltheories}, \cite{Mertens:2020hbs, Mertens:2019tcm,Mertens:2018fds, Blommaert:2018oro}, and therefore one expects various interesting connections among their partition functions and in general, various  ``universalities" between these partition functions.

Two canonical works about Schwarzian limit and its partition function are \cite{Mertens:2018fds,Blommaert:2018oro}, where they discuss precisely the relations among three-dimensional gravity/Chern-Simons theory, Liouville/WZW theory, JT/BF theory and the Schwarzian/particle-on-a-group descriptions.

The main paper which discussed the origin of JT from dimension reduction from $4d$ to $2d$ is \cite{Maldacena:1998uz}, while the low $d$ gravity model of JT has been discussed in \cite{Jackiw:1984je} and \cite{Teitelboim:1983ux}. The connections between their partition functions have been studied in \cite{Ghodrati:2022hbb}.

A very interesting universality is actually the presence, and the dominance, of the Schwarzian modes in the nearly $\text{AdS}_2$ ``tail" or in JT gravity models, where the density of states is $\rho(E) \sim \sinh(2\pi \sqrt{E/\Delta})$, leading to $\rho (E) \sim e^{2\pi \sqrt{E/\Delta} }$ when $E \gg \Delta$, and $\Delta$ here is the energy gap. This leads to a universal contribution to the partition function as 
\begin{gather}
Z_{\text{Schwarzian}} (\beta) \sim \left ( \frac{2\pi}{\beta \Delta} \right)^{3/2} e^{\frac{2\pi^2}{\beta \Delta}},
\end{gather}
which one can detect its presence in many other low-dimensional quantum gravity models' partition functions.

\begin{figure}[ht!]   
\begin{center}
\includegraphics[width=0.8\textwidth]{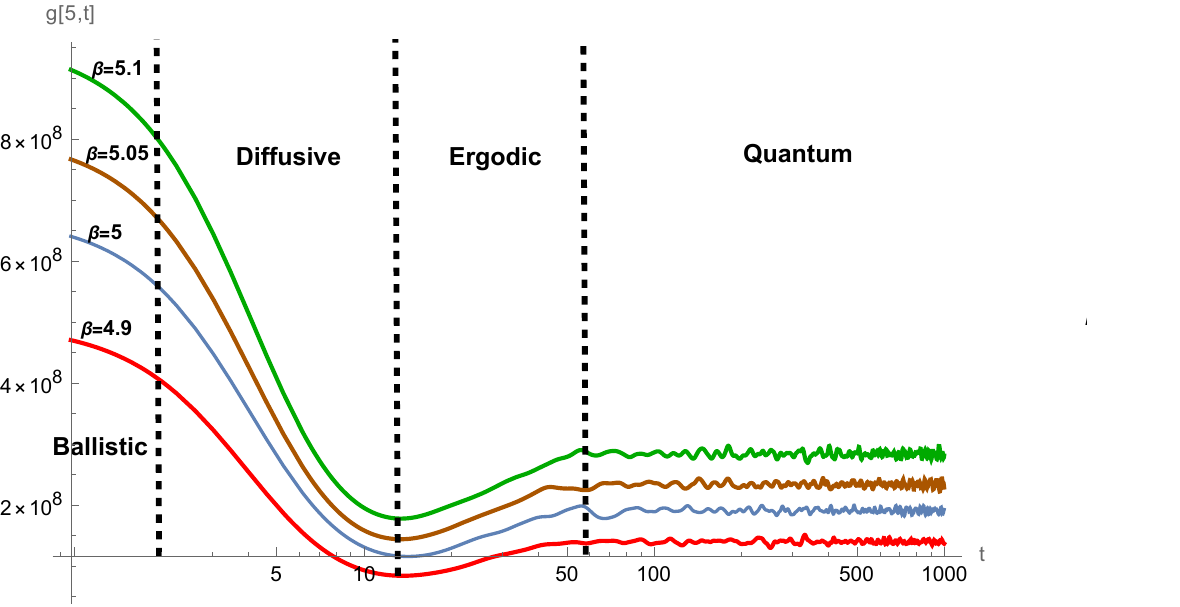} 
\caption{The behavior of the spectral form factor in a Sachdev-Ye-Kitaev (SYK) model, for a sample with $N=1500$ and $q=2$, and with a Hermitian matrix with the size of $N=50$, for different temperatures, i.e., $\beta$, showing different phases during the evolution. The universal behavior of this form factor in chaotic low-d gravitational systems with ramp and plateau can be detected. The four regimes of behavior are present in all of these chaotic systems. Increasing temperature would also decrease spectral form factor (SFF) in all cases.}
\label{fig:SFFphases}
\end{center}
\end{figure}

In the low energy limits $E \to 0$, (related to Sachdev-Ye-Kitaev model universality), one could see that the partition function on a torus or higher-genus models, would factorize into a sum over ``trumpet" geometries where the leading contribution for the genus-0 partition function at low temperature or large $\beta$ would be the JT partition function as 
\begin{gather}
Z_{JT} \approx \frac{e^{S_0}}{4\pi^{3/2}} \left ( \frac{2\pi}{\beta} \right)^{3/2} e^{\frac{2\pi^2}{\beta}}, \ \ \ \text{here} \ \Delta=1,
\end{gather}
which matches the universal late-time spectral form factor of any chaotic system with a ramp and plateau as in Figure \ref{fig:SFFphases}.

Note that this partition function has been found by taking the path integral \cite{Iliesiu:2020obd}
\begin{gather}
Z_{\text{JT}} \lbrack \beta, \phi_r \rbrack = \int \frac{d\mu \lbrack \tau \rbrack }{SL(2, \mathbb{R}) } e^{\int du d\phi_b \text{Sch} (\tau, u)}, \ \ \ \ \ \ d\mu\lbrack \tau \rbrack =  \prod_{\mu \in \lbrack 0, \beta)} \frac{d\tau}{\tau'},
\end{gather}
leading to 
\begin{gather}
Z_{\text{JT}} \lbrack \beta, \phi_r \rbrack \sim \int dE \sinh(2\pi \sqrt{E} ) e^{- \frac{\beta}{2\phi_r}E} \sim \left ( \frac{\phi_b}{\beta} \right)^{\frac{3}{2}} e^{2\pi^2 \frac{\phi_b}{\beta} }.
\end{gather}

Another interesting universal behavior could be found by considering Liouville field theory matter ($c \le 1$), coupled to $2d$ gravity where its partition function is
\begin{gather}
Z_{\text{Liouville}} (A) = \frac{1}{A^3} e^{\frac{A}{2\pi b^2}},
\end{gather}
and $b$ is the Liouville coupling. In the double-scaling limit where $c \to 1$, we get universal critical exponents which are related to the Airy kernel universality in matrix models.

For BF theory, ($B \land F$) in $2d$ as an example of a topological field theory with the action $\frac{k}{4\pi} \int \mathrm{tr} (BF)$, we could see that the partition function is independent of the metric and only depends on topology, where in the large-$k$ limit of $2d$ Yang-Mills we get back to the BF model with a universal leading term ($\sim$ area) which is independent of the detailed measures.

We also have universal Cardy behavior for the growth of states $\rho(E) \sim e^{2\pi }\sqrt{\frac{c}{6} E}$ for $E \to \infty$, and a partition function of $Z(\beta) \sim e^{\frac{\pi^2c}{3\beta}}$, which could be seen in various models such as $\text{AdS}_2$, and many other cases.

For the case of $3d$ Einstein gravity with negative cosmological constant $\Lambda$, where the action is
\begin{gather}
S=\frac{1}{16\pi G} \int d^3x \sqrt{g} (R+ \frac{2}{\ell^2}), 
\end{gather}
the partition function would be
\begin{gather}
Z(\beta, \theta) \sim \exp \left ( \frac{\pi c}{6 \beta} \left ( 1- \frac{\ell^2 \theta^2}{\beta^2} \right ) \right ),  \ \ \ \ \beta \to 0,
\end{gather}
which is a Cardy formula for a dual CFT with the BTZ entropy of $S= \frac{2\pi r_+}{4G}$, which is universal for all UV completions with an $\text{AdS}_3$ boundary condition.

For the case of $S^2$ gravity, which is an Euclidean theory on 2-sphere, we have the action 
\begin{gather}
S= \frac{1}{16\pi G} \int d^2x \sqrt{g} R + \frac{\Lambda}{8\pi G}  \int d^3 x \sqrt{g}.
\end{gather}

As the theory is in $2d$, the $R$-term becomes topological. We then get $\int R = 8\pi$, and the action simplifies to $\to \frac{1}{2G} + \frac{\Lambda A}{4\pi G}$, where the partition function would be $Z= e^{-\frac{1}{2G} - \frac{\Lambda A_0}{4\pi G} }$. The exponential of $- \Lambda A_0/(4 \pi G)$, is universal as it appears in the large-area expansion of any $2d$ quantum gravity with a cosmological constant.

The case of the Wess-Zumino-Witten model on a $2d$ surface with level $k$, coupled to gravity also shows universality, and the partition function is given by the Verlinde formula, as
\begin{gather}
Z_{\text{WZW}}^{(g)} \sim k^{(g-1) \text{dim}(G)} e^{k S_{\text{classical}}},
\end{gather}
where again the leading exponential in $k$ is universal as it only depends on the topology and the central charge $c= \frac{k \text{dim} (G) }{k+h}$, in the large-g expansion.

In the case of wormholes we also have universality, as in $3d$ gravity, the partition function of a genus-g would be a sum over geometries with the leading contribution at large $c$ as $e^{c S_{\text{classical}}}$ which can be computed by Weil-Petersson volume and is related to Airy function in JT by dimension reduction.

In \cite{Boruch:2023bte}, also, the connections between $\mathcal{N}=2$ supersymmetric Sachdev-Ye-Kitaev (SYK) model and supersymmetry has been discussed.  One specific universality they found is the connections between the ground state subsector and the total dimension of SYK Hilbert space where the dual interpretation of this fraction is the probability that the zero temperature wormhole (which is a supersymmetric ER bridge) gets a vanishing length. So one could suggest that the probability for each length of the wormhole corresponds to a specific fraction between sets of state subsectors and total dimension of the SYK Hilbert space, or correspondingly to a specific temperature of the Hartle-Hawking wavefunction.

Moreover, in \cite{Ahmadain:2022gfw}, the microstates and partition functions of $2d$ non-supersymmetric asymptotically flat black holes have been studied. The double scaling limit of the partition functions would lead to a dominant term which corresponds to a bound state and is dual of the black hole microstates.  In addition, in \cite{Betzios:2022pji, Betzios:2025eev}, the connections between $2d$ Sine-Liouville theory and black holes were studied where the authors performed a very thorough analysis of the partition function in terms of representations that offers a complementary and quite technically complete picture with respect to \cite{Ahmadain:2022gfw}. These pictures could help to extract further universalities in these low-dimensional gravity models, and specifically their partition functions.

The main understanding for the universalities to be found in low dimension gravity models can also be understood from the physical intuitions presented in Kitaev’s talk at IAS \cite{Kitaev:2015Caltech} where the bulk has been considered as a filter changing the spectral density at the horizon, to the one at the asymptote, which for the case of $\text{AdS}_2$ geometry would lead to 
\begin{gather}
\text{At the horizon: } \ A_{\psi, \psi}^{\text{hor}} (\omega) = \text{const}, \ \ \ \ \text{At the infinite asymptote} \sim \cosh \frac{\beta \omega}{2} \Big | \Gamma ( \frac{3}{4} - \frac{i \beta \omega}{2\pi} ) \Big |^2.
\end{gather}

Now, this picture of considering the bulk in any low-dimensional gravity model as lens changing the spectral density with a specific universal behavior can be extended to other gravity models which have holographic duals as well.

The motivation of this work is indeed to find more of such universalities and connections between the partition functions of various $2d$ gravity models and their corresponding higher-dimensional counterparts. Specifically, we would like to further investigate the connections between the partition functions of JT and supersymmetric models in $\text{AdS}_2$ and $S^2$. We also want to check the connections between the partition function of $3d$ BTZ and $\text{AdS}_3$ and the dimension reduced $2d$ JT. In addition, the corresponding gauged case of $3d$ Chern-Simons and $2d$ BF models would also be checked.

\section{First observation: The connections between partition function of JT and low-dimensional supersymmetric models}

In \cite{GonzalezLezcano:2023cuh}, the global Euclidean $\text{AdS}_2$ on $\mathcal{N}=(2,2)$ and on $S^2$ have been calculated, and the effects of the zero modes and their superpartners have also been calculated where they found that the resulting partition function would depend on the total size of the manifold, which could be either $\text{AdS}_2$ or $S^2$. 

In this section we would like to compare the partition function that they found for $R$-symmetric $\mathcal{N}=(2,2)$ Abelian theory on $S^2$ and $\text{AdS}_2$ using localization technique, and the partition function of JT.

The Euclidean action of JT gravity with an end-of-the-world brane (EOW brane) behind the black hole horizon can be written as 
\begin{gather}
I = I_{JT} + \mu \int_{\text{brane}} \mathrm{d} s,
\end{gather}
where this integral is along the worldline of the EOW brane, which then can be compared with the case of other $2d$ gravity models.

The pure JT action is 
\begin{gather}
I_{JT}= - \frac{S_0}{2\pi} \left \lbrack \frac{1}{2} \int_{\mathcal{M}} \sqrt{g} R + \int_{\partial \mathcal{M}} \sqrt{h} K\right \rbrack - \left \lbrack \frac{1}{2} \int_{\mathcal{M}} \sqrt{g} \phi ( R+2) + \int_{\partial \mathcal{M}} \sqrt{h} \phi K \right \rbrack. 
\end{gather}

The partition function of JT has been studied in various works such as \cite{Yang:2018gdb, Kitaev:2018wpr, Penington:2019kki}, which can be written as
\begin{gather}
Z_n= e^{S_0} \int_0^\infty \mathrm{d} \ell_1 . . .  \  \mathrm{d} \ell_{2n} e^{\frac{\ell_1 + . . . + \ell_{2n} }{2} } I_{2n} (\ell_1, . . . \ell_{2n}) \varphi_\beta(\ell_1) e^{-\mu \ell_2} . . . . \varphi_\beta ( \ell_{2n-1}) e^{-\mu \ell_{2n} }\nonumber\\
= \int_0^\infty \mathrm{d} s \rho(s) y(s)^n; \ \ \ \ \ \ \ y(s)= e^{- \frac{\beta s^2}{2} }   2^{1-2 \mu} \big | \Gamma( \mu- \frac{1}{2}+is) \big |^2,
\end{gather}
which is in the form of 
\begin{gather}
Z_n = I_{2n} + \varphi \times n.
\end{gather}

In the above relation, the object $\varphi$ is the Hartle-Hawking state in the geodesic basis, where it computes the path integral from the asymptotic boundary (which is related to the renormalized length $\beta$) with the geodesic of regularized length $\ell$.

So the $\varphi$ part of JT partition function is related to the regularized part of $\text{AdS}_2$ and $S^2$ in the above two partition functions, namely $ \left (\frac{L}{L_0} \right)^{-i \sigma_1 L}$ in the case of $\text{AdS}_2$ and  $\left ( \frac{L}{L_0} \right )^{- 2 i \sigma_0 L}$ in the case of $S^2$.

Also, the expression for the functions $\varphi_\beta (\ell)$ is
 \begin{gather}
 \varphi_\beta (\ell)= 4 e^{- \frac{\ell}{2} } \int_0^\infty \mathrm{d} s \rho (s) e^{\frac{- \beta s^2}{2} } K_{2i s} ( 4 e^{- \frac{\ell}{2} } ).  
 \end{gather}

The object $I_{2n} ( \ell_1, . . ., \ell_{2n})$ in the JT gravity path integral has $2n$ geodesic boundaries with the fixed regularized lengths $\ell$ and has the expression as below
 \begin{gather}
 I_{2n}(\ell_1, . . . . , \ell_{2n}) = 2^{2n} \int_0^\infty \mathrm{d}s \rho(s) K_{2is} ( 4 e^{- \frac{\ell_1}{2} }) . . . . K_{2is} (4 e^{- \frac{\ell_{2n} }{2} } ); \ \ \ \ \rho(s)= \frac{s}{2 \pi^2} \sinh(2\pi s). 
 \end{gather}

Actually, the boundary of $I_{2n}$ has $n$ geodesics which correspond to the EOW branes, and another $n$ geodesics that are glued to Hartle-Hawking states.

Now we can compare the partition function of JT gravity with other interesting low-dimensional gravity models. Specifically, the partition functions of $R$-symmetric $\mathcal{N}= (2,2)$ theory on $S^2$ and $\text{AdS}_2$, which have been discussed in \cite{GonzalezLezcano:2023cuh}, can be compared with the JT partition function to detect similarities and therefore universalities. Note that initially the $S^2$ partition function was first calculated in \cite{Doroud:2012xw} for the A-type, and in \cite{Doroud:2013pka} for the B-type, and in \cite{Gomis:2012wy}, the physical interpretations were discussed.

The total action that the authors of \cite{GonzalezLezcano:2023cuh} considered is
\begin{gather}
S_{\text{tot}} = S_{\text{v.m.}}+S_{\text{ghost}} +S_{\text{FI}}+S_{\text{top.}}+S_{\text{c.m.}}, 
\end{gather}
where the vector multiplet part can be written as
\begin{gather}
S_{\text{v.m.}} = \frac{1}{g^2_{\text{YM}}} \int \mathrm{d}^2 x \sqrt{g} \left \lbrack \mathcal{L}_{\text{v.m.}}^{\text{bulk} } +D_{\mu} V^\mu_{\text{v.m.}} \right \rbrack,
\end{gather}
and $g_{\text{YM}}$ is the super-renormalizable gauge coupling. One expects that $g_{\text{YM}}$ would be related to $S_0$, which is the zero-temperature entropy of the eternal $2d$ black hole.

For $S^2$, the supersymmetric Fayet-Iliopoulos (FI) parameter and the topological term are
\begin{gather}
S_{\text{FI}}^{\text{S}^2}+S_{\text{top}}^{\text{S}^2}=-i \xi \int \mathrm{d}^2x \sqrt{g}\hat{D} +i \frac{\vartheta}{2\pi} \int \mathrm{d}^2 x \sqrt{g} F_{12},
\end{gather}
and for the case of $\text{AdS}_2$ it is
\begin{gather}
S_{\text{top.}}^{\text{AdS}_2} = i \frac{\vartheta}{2\pi} \int \mathrm{d}^2 x \sqrt{g} F_{12} + i \frac{\vartheta}{2\pi} \int \mathrm{d} \theta (\bar{\epsilon} \epsilon \sigma + i \bar{\epsilon} \gamma_3 \epsilon \rho).
\end{gather}

So, again, as one would expect, the topological parameter $\vartheta$ in the $R$-symmetric $\mathcal{N}= (2,2)$ theory is related to the topological term $S_0$ in JT gravity.

The metric of the $S^2$ and $\text{AdS}_2$ backgrounds are
\begin{gather}
S^2: \ \ \ \ \ \ \ ds^2= L^2 (d\psi^2 + \sin^2 \psi d\theta^2), \ \ \ \ \ \ 0\le \psi < \pi, \ \ 0 \le \theta < 2\pi, \nonumber\\
\text{AdS}_2: \ \ \ \ ds^2 = L^2 ( d\eta^2 + \sinh^2 \eta d\theta^2 ), \ \ \ \ \ \ 0 \le \eta < \infty, \ \ 0 \le \theta < 2\pi, 
\end{gather}
where $L$ is the size of the background. 

The finite-dimensional localization measure for these two backgrounds are
\begin{gather}
\text{for} \ S^2 : \sum_{\mathbf{m} \in \mathbb{Z} } \int \frac{\mathrm{d}( \sigma_0 L) }{(g_{\text{YM}} L_0)^2},  \ \ \ \ \ \ \ \  \ \ \ \ \ \ \ \text{for} \ \text{AdS}_2: \ \int \frac{\mathrm{d} (\sigma_1 L)  }{g_{\text{YM} } L_0}. 
\end{gather}
 
The partition function for $S^2$ is
\begin{gather}
Z_{S^2}= \left ( \frac{L}{L_0}   \right )^{1-r} \sum_{\mathbf{m}= - \infty }^ \infty e^{- i \vartheta \mathbf{m} } \int \frac{\mathrm{d}( \sigma_0 L) }{(g_{\text{YM}} L_0)^2}  \left ( \frac{L}{L_0} \right )^{- 2 i \sigma_0 L} e^{4 \pi i \xi \sigma_0 L} \frac{\Gamma (\frac{r}{2} + i \sigma_0 L - \frac{\mathbf{m} }{2} ) }{\Gamma (1- \frac{r}{2} - i \sigma_0 L - \frac{\mathbf{m} }{2} )}.
\end{gather}

As discussed in \cite{GonzalezLezcano:2023cuh}, the parameter $\sigma_0 L$ could then be absorbed into the renormalized FI parameter as $\xi_{\text{ren}} = \xi - \frac{1}{2\pi} \log \left ( \frac{L}{L_0} \right )$, and therefore it is related to the renormalization parameter in $2d$ gravities.

Now we can compare the above two partition functions with the JT partition function of \cite{Penington:2019kki}.

Also, it should be emphasized here that $g_{\text{YM}}$ is not a running coupling and therefore the contribution from $(g_{\text{YM} } L_0 )^{-2}$ is just a purely irrelevant numerical factor. Also, this partition function includes the scale-dependent factor $-1+(1-r)$, which is related to the scaling anomaly, and an additional $``1"$ which comes from the zero modes of the ghost fields.

The partition function of the $\text{AdS}_2$ part would then become
\begin{gather}
Z_{\text{AdS}_2} = \left (\frac{L}{L_0} \right )^{1- \frac{r}{2} + \rho_0 L } \int \frac{\mathrm{d} (\sigma_1 L) }{g_{\text{YM}} L_0} \left (\frac{L}{L_0} \right)^{-i \sigma_1 L} \frac{1}{\Gamma (1- \frac{r}{2}- i \sigma_1 L + \rho_0 L ) } . 
\end{gather} 

Similar to the previous case, the above partition function contains a contribution $-\frac{1}{2}+ \frac{1}{2} (1-r)$ from the anomaly, and the ``$1$" which comes from the boundary zero modes. In addition, the contribution from $\rho_0 L$ can be absorbed into the $R$-charge.

So the main result here is that, in the above two partition functions, the term $e^{4\pi i \xi \sigma_0 L}$ is related to the Hartle-Hawking state, and the Gamma function factors $ \frac{\Gamma (\frac{r}{2} + i \sigma_0 L - \frac{\mathbf{m} }{2} ) }{\Gamma (1- \frac{r}{2} - i \sigma_0 L - \frac{\mathbf{m} }{2} )}$ in the case of $S^2$, and $\frac{1}{\Gamma (1- \frac{r}{2}- i \sigma_1 L + \rho_0 L ) } $ in the case of $\text{AdS}_2$, are related to the brane states. Also, the parameter $s$ corresponds to the energy $s^2/2$ of a particular asymptotic region, which is related to $\sigma_0 L$ in the case of $S^2$ and $\sigma_1 L$ in the case of $\text{AdS}_2$.

Note that one main result of \cite{GonzalezLezcano:2023cuh}, is that their partition function on $S^2$ or $\text{AdS}_2$ can be written in the form of $Z \sim \text{(local anomaly term)} \times \text{(global zero-mode term)}$. So, based on the universality behavior we saw in low-dimensional gravity models, we would expect that the JT partition function could be written in such a way as well. Particularly, the term containing the gravitational ``density of states" $\rho(s)$, could be imagined as the term describing the ``local anomaly", and the the term containing $y(s)^n$, could be imagined as the one describing the global zero-mode term.

Specifically, in \cite{Yang:2018gdb}, it has been shown that $\rho(s)$ contains term with residue gauge $\frac{1}{2\pi}$, and a term which describes a charged particle, ``confined" in a magnetic field, therefore emphasizing the ``local" behavior of such term. The reason that the second term is related to zero-modes is because the Bessel function could actually be represented as the fixed energy ``microstate" $\ket{\mathcal{E}}$, where the wavefunction is $\Psi (E; \ell)=\rho(E) \frac{4}{\ell} K_{i \sqrt{8E}} (\frac{4}{\ell})$.

Similarly, in the case of \cite{GonzalezLezcano:2023cuh}, zero modes are constant modes of ghost and anti-ghost fields, behaving like the main propagators in JT.

Also, note that for the case of $\text{AdS}_2$, unlike $S^2$, one has ``boundary zero modes", which are pure gauge modes, with non-normalizable parameter $\Lambda_{\text{bdry}}^{\ell}$ \cite{Camporesi:1992tm}, which can be written as 
\begin{gather}
A^{(\ell)}_{\text{bdry}} = d\Lambda_{\text{bdry}}^{(\ell)}, \ \ \ \ \ \Lambda_{\text{bdry}}^{(\ell)} = \frac{1}{\sqrt{2\pi |\ell |}} \left(\frac{\sinh \eta}{1+\cosh \eta}\right )^{|\ell |} e^{i \ell \theta}, \ \ \ \ \ell= \pm 1, \pm 2, ... .
\end{gather}

Now, this can be compared with the propagator of a non-relativistic particle in a magnetic field as \cite{Comtet:1986ki}, 
\begin{flalign}
G(u, \boldsymbol{x} _1, \boldsymbol{x}_2) &= e^{ i \varphi (\boldsymbol{x} _1, \boldsymbol{x}_2)} \int_0^\infty ds s e^{-u \frac{s^2}{2} } \frac{\sinh 2 \pi s}{2\pi (\cosh 2 \pi s + \cos 2 \pi b) } \frac{1}{d^{1+ 2 i s} }\times \nonumber\\ &
\times \  {}_2F_1( \frac{1}{2} - b + i s, \frac{1}{2}+ b + is, 1, 1- \frac{1}{d^2} ).\nonumber\\
d &= \sqrt{\frac{(x_1 - x_2)^2 + (y_1 + y_2)^2}{4 y_1 y_2} },\nonumber\\
e^{i \varphi (\boldsymbol{x} _1, \boldsymbol{x}_2) } &= e^{- 2 i b \arctan \frac{x_1 -x_2}{y_1 + y_2} }.  
\end{flalign}
which, as shown in \cite{Penington:2019kki}, was used to derive the partition function of JT gravity. As mentioned before, the canonical partition function of a quantum mechanical system can be written as 
\begin{gather}
Z_{\text{particle}} = V_{AdS} \int_0^\infty ds e^{- \beta \frac{s^2}{2}} \frac{s}{2\pi} \frac{\sinh (2\pi s)}{\cosh(2\pi q) + \cosh(2\pi s)}.
\end{gather}

Therefore, one could expect that the partition function of these supersymmetric low-dimensional gravity models can also be written using the picture of charged particles in magnetic fields. This is yet a conjecture as unlike the JT case, the $\mathcal{N}=(2,2)$ theories are defined on fixed backgrounds, and a precise map between localization parameters and particle variables has not yet been established. However, we propose our tentative maps here and leave a rigorous derivation for future works.

Also, in the case of supersymmetric model, the cohomological variables $(\Phi, Q_{eq} \Phi, \Psi, Q_{eq} \Psi)$, where $\Phi$ is the elementary boson and $\Psi$ is the elementary fermion, and $Q_{eq}\Phi$, $Q_{eq}\Psi$ are their superpartners, are related to the Bessel functions $K_{2is}$ in the JT partition function, which are related to fixed-energy propagators.

Note that the Bessel functions in the JT partition function are well-behaved at short distance, since
\begin{gather}
K_{2is} ( \frac{4}{\ell}) \simeq \sqrt{\frac{\pi}{8\ell} } e^{-\frac{4}{\ell}}, \ \ \ \ \ell= \frac{|x_1 - x_2 |}{\sqrt{z_1 z_2}} \to 0, 
\end{gather}
which is compatible with the our picture of their connections with the cohomological variables. 
Also, as one expect from particle-like behavior in JT, the special functions 
\begin{gather}
f_{k,s}(x,z)= \sqrt{z} e^{ikx} K_{2is} (2\sqrt{2ikz}),
\end{gather}
would be delta-function normalizable eigenmodes of large-$q$ Hamiltonian, again confirming the relation to global zero-mode terms and super-particles, as it is singular at short times and distances.
\begin{gather}
\int \frac{dx dz}{z^2} f_{k_1,s_1} f_{k_2, s_2} = \delta(k_1 - k_2) \delta(s_1-s_2) \frac{\pi^3}{2s \sinh(2\pi s)}, \nonumber\\
\tilde{K}(u, \mathbf{x_1}, \mathbf{x_2}) \sim \delta(x_1 - x_2 + uz_2) e^{- \frac{(z_1 - z_2)^2}{2u z_2} }.
\end{gather}

For the supersymmetric localization saddle, we can perform the integration along the locus as
\begin{gather}
\mathcal{M}_{\text{loc}} = \{ \varphi_{\text{loc}} \big | Q_{eq} \mathcal{V}(\varphi_{\text{loc}}) \big |_{\text{boson}} =0,  \  \chi=0 \},
\end{gather}
and we get
\begin{gather}
Z= \lim_{t \to \infty} Z_t = \int_{\mathcal{M}_{\text{loc}} } \mathcal{D} \varphi_{\text{loc}} e^{-S(\varphi_{\text{loc}} )} {Z'}_{1-\text{loop}}^{Q_{eq}}.
\end{gather}

For the case of JT, the functions $I_{2n}$ and $\varphi_\beta$ are related to this 1-loop determinant of the quadratic kinetic operator.

For the case of $\text{AdS}_2$, the one-loop part of partition function is \cite{GonzalezLezcano:2023cuh}
\begin{gather}
Z_{1-\text{loop}} = \left( \frac{L}{L_0}\right)^{\frac{1}{2} (1-r-2i \sigma_1 L + 2 \rho_0 L) }\Gamma \left (\frac{r}{2}- i \sigma_1 L + \rho_0 L \right),
\end{gather}
so, for JT, the corresponding term would be $ \sim \rho(s) \left(2^{1-2\mu} | \Gamma(\mu- \frac{1}{2} + is) |^2\right)^n $.

We could also check the recently found partition functions, such as the one for the case of $\mathcal{N}=(2,2)$ theories on spindle with twist and anti-twist, which have two opposite $R$-charges, as in \cite{Jeon:2026kxn}, the partition function has been found as
\begin{gather}
Z^\eta_{\mathbb{\sum}}= \frac{L}{L_0} e^{- \frac{r}{2} (\frac{2\pi \xi - i \eta \theta}{n_1} + \frac{2\pi \xi + i \theta}{n_2} ) } e^{-i (1+ \eta) \frac{L}{L_0} e^{-2 \pi \xi}  \sin \theta} \nonumber\\
\times \left(\frac{1-e^{-(2\pi\xi - i \eta \theta)} }{1- e^{-(2\pi \xi - i \eta \theta)/n_1} } \right) \left (\frac{1- e^{-(2\pi \xi+ i \theta) } }{1- e^{- (2\pi \xi + i \theta)/n_2 } }\right).
\end{gather}

The above partition function also can be imagined as two particles in a magnetic field, similar to the picture in \cite{Yang:2018gdb}. However, instead of only one particle, here in the spindle case, the ratio between the circles of the paths of particles, the ratio of their energies and fluxes, and the ratio of spindle size relative to a reference length scale are important. So, here $L/L_0$ is related to $e^{S_0}$, and we also have the relations
\begin{gather}
\frac{L}{L_0}(\text{SFT on spindle})  \to e^{S_0} (\text{JT}), \nonumber\\
 e^{- \frac{r}{2} (\frac{2\pi \xi - i \eta \theta}{n_1} + \frac{2\pi \xi+i \theta}{n_2} ) } (\text{SFT on spindle}) \to e^{\frac{\ell_1 + ... + \ell_{2n} }{2} } (\text{JT}), \nonumber\\
 e^{- \mu \ell_2}....e^{- \mu \ell_{2n}} (\text{SFT on spindle})  \to e^{-i(1+\eta) \frac{L}{L_0} e^{-2\pi \xi} \sin \theta } (\text{JT}), \nonumber\\
 I_{2n} (\ell_1, ..., \ell_{2n}) (\text{SFT on spindle}) \to \left(\frac{1-e^{-(2\pi\xi - i \eta \theta)} }{1- e^{-(2\pi \xi - i \eta \theta)/n_1} } \right) \left (\frac{1- e^{-(2\pi \xi+ i \theta) } }{1- e^{- (2\pi \xi + i \theta)/n_2 } }\right) ((\text{JT})).
\end{gather}

Specifically, note that in this case, the $R$-symmetry flux should satisfy the following quantization condition
\begin{gather}
\frac{1}{2\pi} \int_{\sum} dA^R = \frac{1}{2} \ . \ \frac{\eta n_2 - n_1}{n_1 n_2} = \eta \frac{\chi - \eta}{2},
\end{gather}
which is related to $- \beta \frac{s^2}{2}$. Note that the relation $\eta= -1$ is related to twist and $\eta=1$ to anti-twist, and $\chi$ is the Euler characteristic of the spindle, and they determine the ``topological twist" properties. One could imagine that when the spindle parameters are equal, $n_1=n_2$, the results become more similar to the partition function of JT, which is indeed the case.

Also, based on the pattern of the universalities we see in the partition function of many quantum gravity models, we expects that when $n_{1,2} \to 1$, one reaches another well-known result, which would be actually the chiral multiplet one-loop determinant on the $\Omega$-deformed sphere $S_{\Omega}^2$ in \cite{Closset:2015rna}, as 
\begin{gather}
\lim_{n_1, n_2 \to 1}  \left(\frac{L}{L_0} \right)^{\mathfrak{b}} \frac{\Gamma(\frac{1- \mathfrak{b} - \mathfrak{c}}{2} + \frac{r}{4} \chi_- + q_G \gamma_G) }{\Gamma(\frac{\mathfrak{b}-\mathfrak{c}+1}{2} + \frac{r}{4} \chi_- + q_G \gamma_G)} = \left( \frac{L}{L_0} \right)^{1-r+\mathfrak{m}} \frac{\Gamma (\frac{r - \mathfrak{m}}{2} + \gamma_G) }{\Gamma( 1- \frac{r- \mathfrak{m}}{2} + \gamma_G)}.
\end{gather}

For finding further connections between various $2d$ gravity models, one could note works such as \cite{Joung:2023doq}, where the effects of edge modes of JT and BF models have been investigated.

In \cite{Joung:2023doq}, the authors introduced wiggling boundaries and the would-be gauge modes, building the path to find universalities, such as the relation between the Haar measure of the Gauss decomposition and Iwasawa decomposition as
\begin{gather}
\int \frac{\mathcal{D} u \ \mathcal{D}y\  \mathcal{D}f}{\prod_{\theta} \lbrack y (\theta) \rbrack^2} = \int \frac{\mathcal{D}E\  \mathcal{D} \Lambda \ \mathcal{D}F}{\prod_\theta \lbrack \Lambda(\theta) \rbrack^2}.
\end{gather}

This relation could help to find connections between their quantum partition functions. However, this is an important problem that remains to be addressed to find the transformation properties of the quantum gravity partition function, which \cite{Joung:2023doq} did not address.

 Also, in \cite{Ecker:2023sua}, the equivalences between various $2d$ dilaton gravities, their dual holographic pictures, and their corresponding asymptotic symmetries have been examined, which could further clarify the connections between the partition functions among such $2d$ gravity models.

\subsection{Connections between partition functions in various dimensions}

As previously discussed in section \ref{sec:intro}, and as can be seen from Figure \ref{fig:dimHoltheories}, an interesting picture in low-dimensional gravity is that various limits of a specific partition function could lead to the partition functions in other gravity models or the partition functions in other dimensions.

For example, the Cardy-like limit involves shrinking the circle of $S^1$ with the limit of $\beta \to 0$ in two dimensions and $R_{S^1}/R_{S^d} \to 0 $ in higher dimensions, which could map the partition functions in $3d$ to $2d$ case. These cross-dimensional limits and Kaluza-Klein dimension reductions of various models, and the connections between the resulting partition functions in various dimensions, have been recently investigated in \cite{Amariti:2023ygn}, where they also found that the R-symmetry or flavor symmetry two-point current correlation function would determine the universal behavior of the partition function of $3d$ large-$N$ superconformal field theories, leading to a Cardy-like formula.

Specifically, in that work, similar to our observation in the previous section, it has been shown that there is a universal behavior in the ``free energy" of $3d$ SCFTs (superconformal field theories) on $S_\beta^1 \times S^2$, which is related to the coefficients of R-symmetry or flavor symmetry as 
\begin{gather}
\text{Re} ( F_{S^1 \times S^2}) = - \frac{1}{2} ( m \bar{L} )^2 \tau_{aa} + \mathcal{O} ( \beta ), \ \ \ \ \ \ \  \text{Im} (F_{S^1 \times S^2}) \propto - \frac{1}{2} (m \bar{L})^2 \kappa_{aa},
\end{gather}
 where $m$ is the scalar in the vector multiplet and $\bar{L}$ is the length scale of $3d$ curved background.
 
The $3d$ partition in their work has been found as
\begin{gather}
Z_{\mathcal{M}_{g,p}} (\nu) = \sum_{\hat{u} \in \mathcal{S}_{BE} }\mathcal{F}(\hat{u}, \nu)^p \ \mathcal{H}(\hat{u}, \nu)^{g-1} \prod_{i} \Pi_i (\hat{u}, \nu)^{n_i} = \sum_{\mathfrak{m} \in \mathbb{Z}} \int_{\mathcal{C}}du \ \mathcal{J} (u) \ \Pi(u)^{\mathfrak{m}} ,
\end{gather}
where
\begin{gather}
\mathcal{H}(u) \equiv \text{det}_{ab} \left ( \frac{\partial^2 \mathcal{W}(u) }{\partial  u_a \partial u_b} \right)  \ . \  e^{2\pi i \tilde{\Omega}(u) },\nonumber\\
\mathcal{F}(u) \equiv \text{exp} \left ( 2\pi i \left (  \mathcal{W}(u) - u_a \frac{\partial \mathcal{W} (u) }{\partial u_a} \right) \right ),\nonumber\\
\Pi_i(u, \nu) \equiv \text{exp} \left ( 2\pi i \frac{\partial \mathcal{W}(u , \nu) }{\partial \nu_i} \right ).
\end{gather}

Then, upon dimension reduction, one could connect to them to the JT partition function. One could see that the product of the fluxes $\mathcal{F}(\hat{u}, \nu)^p \mathcal{H}(\hat{u}, \nu)^{g-1} $ is related to the density matrix in JT and the flux $\Pi_i$ to the functions $g(s)$, and $\Gamma$ function in the JT partition function. Note that these $\Gamma$ function are also some form of a geometric series. One could specify in more details the contributions of each part.

To further analyze these universalities and the connections with the JT gravity, we could investigate the structure of JT more explicitly and connect it with those universalities mentioned above. Specifically, the integrand in JT gravity is in the following form
\begin{gather}
I_{\text{JT}}=\frac{s}{2\pi^2} \sinh (2 \pi s)  \left ( e^{- \frac{\beta s^2}{2} } 2^{1- 2 \mu} \big | \Gamma ( \mu - \frac{1}{2} + i s) \big |^2  \right )^n.
\end{gather}

\begin{figure}[ht!]   
\begin{center}
\includegraphics[width=0.45\textwidth]{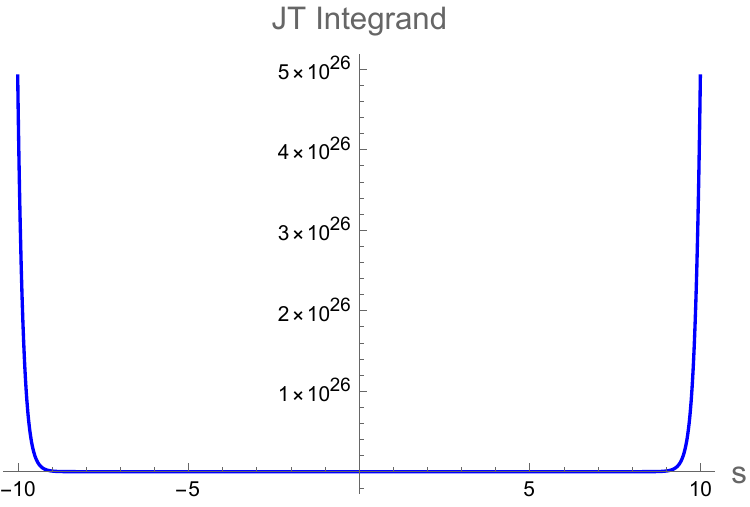}  
\includegraphics[width=0.5\textwidth]{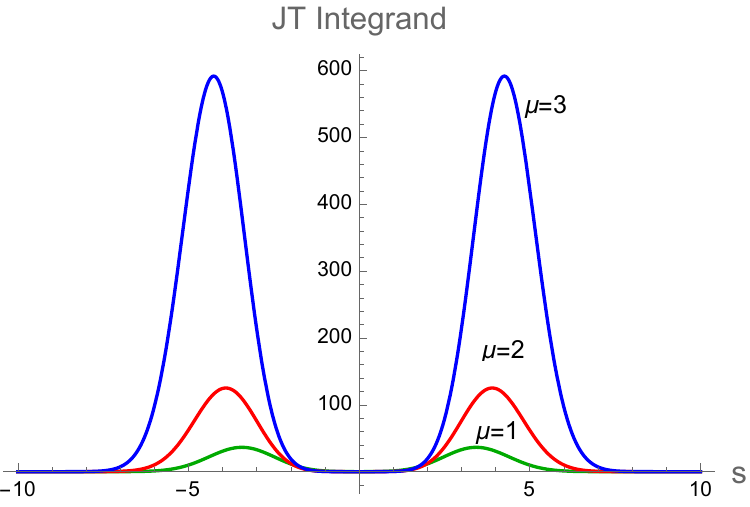}  
\caption{The behavior of the JT integrand is shown for $n=0$, in the left section, and $n=1$ in the right section. In this figure, $n$ is the number of physical boundaries.}
\label{fig:JTInt1}
\end{center}
\end{figure}

We can see its behavior in Figure \ref{fig:JTInt1}.  Here, $n$ is the number of physical boundaries. It can be seen that by increasing ``$n>0$", the general behavior of the function remains similar to the case of $n=1$, but the maximum value becomes very close to zero. That is why higher $n$ is strongly suppressed and also the behavior of the partition function in this limit would be $ Z_n\sim e^{S_0 \chi}$.

\begin{figure}[ht!]   
\begin{center}
\includegraphics[width=0.45\textwidth]{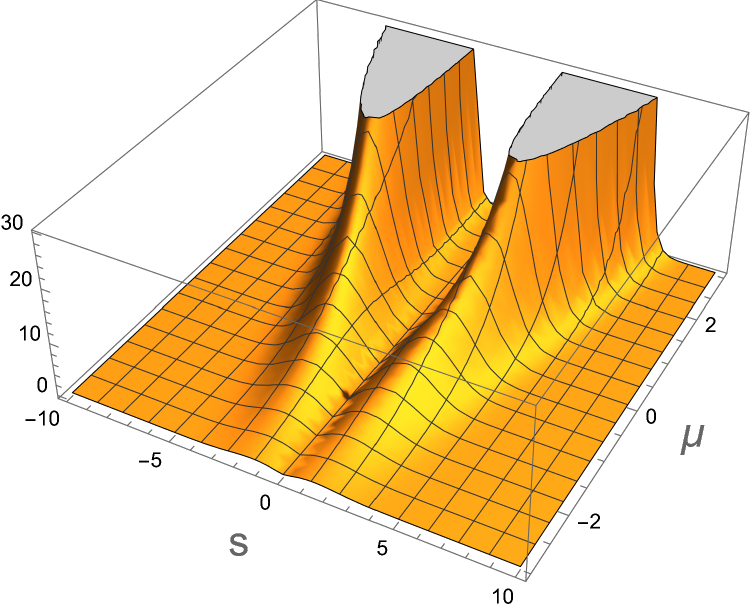}  \ \ \ \ \ \ \ 
\includegraphics[width=0.45\textwidth]{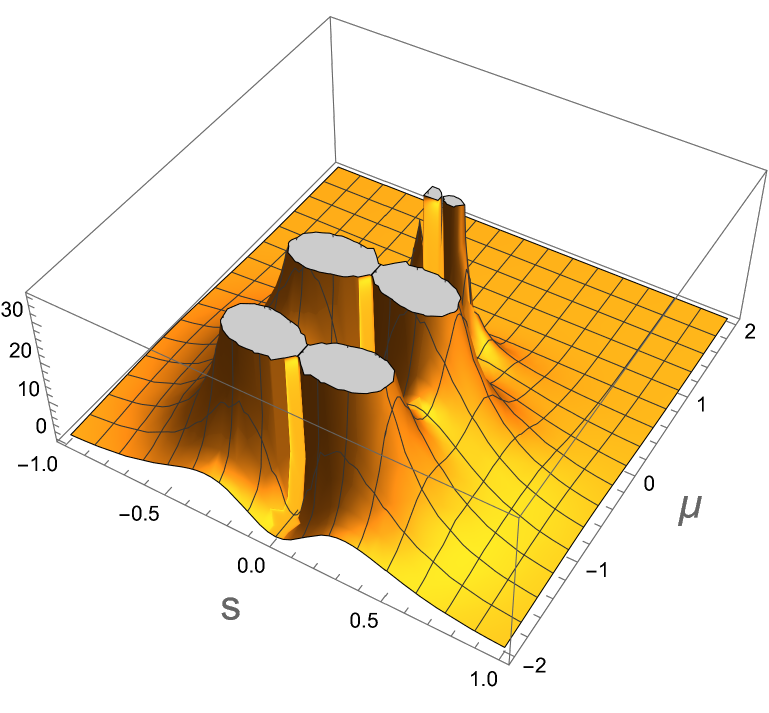}  
\caption{The behavior of JT integrand versus $s$ and $\mu$ is shown, for the case of $n=1$, in the left panel, and $n=2$ in the right panel.}
\label{fig:JT3dIntegrand}
\end{center}
\end{figure}

In Figure \ref{fig:JT3dIntegrand}, the behavior of JT integrand versus $s$ and $\mu$ is shown. When $n=2$, three main peaks would be generated, for positive $s$ and negative $s$, which are symmetric, and the general behavior would remain similar by increasing $n$ to higher values.

\begin{figure}[ht!]   
\begin{center}
\includegraphics[width=0.45\textwidth]{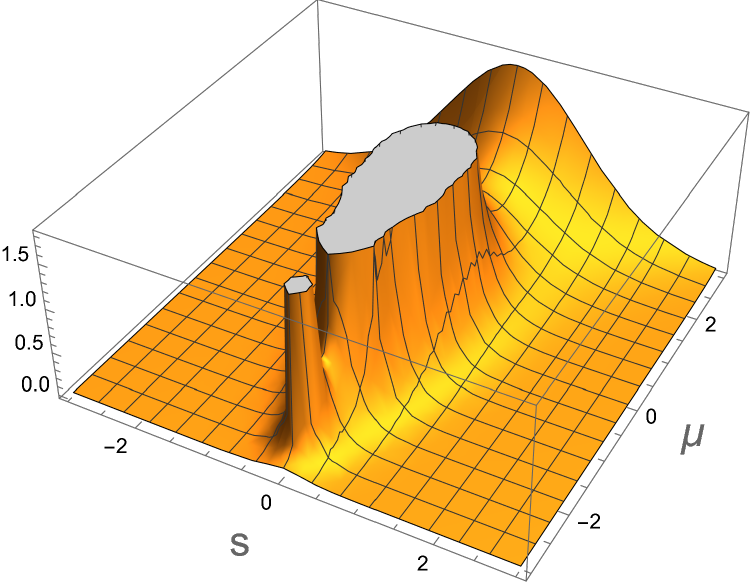}    \ \ \ \ \ \
\includegraphics[width=0.45\textwidth]{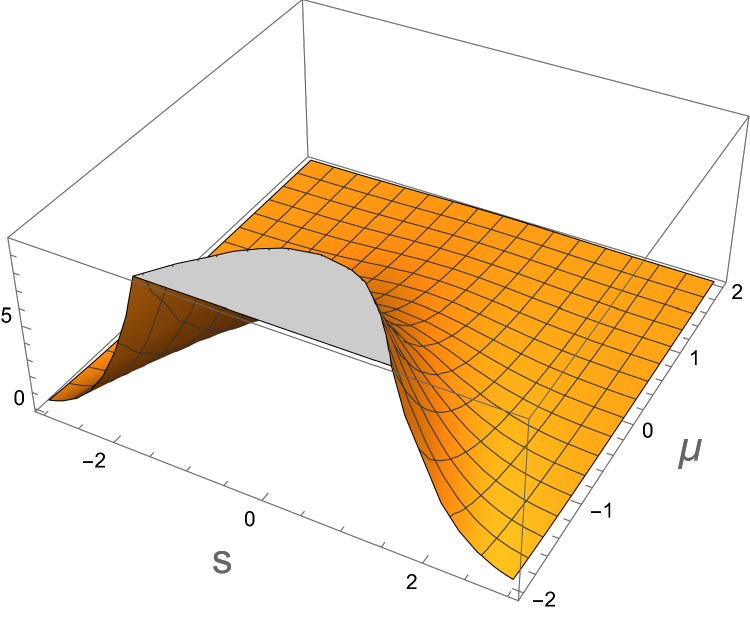}  
\caption{The behavior of the function $\big | \Gamma ( \mu - \frac{1}{2} + i s ) \big |^2 $ corresponding to the EOW brane state is shown on the left panel, and the behavior of the function $ e^{-\frac{\beta s^2}{2} } 2^{1- 2 \mu} $ corresponding to the Harte-Hawking state is shown on the right panel.}
\label{fig:JT2partPartition}
\end{center}
\end{figure}

In Figure \ref{fig:JT2partPartition}, the behavior of the function $\big | \Gamma ( \mu - \frac{1}{2} + i s ) \big |^2 $ which corresponds to the EOW brane state and the behavior of the function $ e^{-\frac{\beta s^2}{2} } 2^{1- 2 \mu}$ which corresponds to the Harte-Hawking state is shown on the right panel. Most of the partition functions in low-dimensional quantum gravity models could be considered a combination of these two types of functions.

\begin{figure}[ht!]   
\begin{center}
\includegraphics[width=0.45\textwidth]{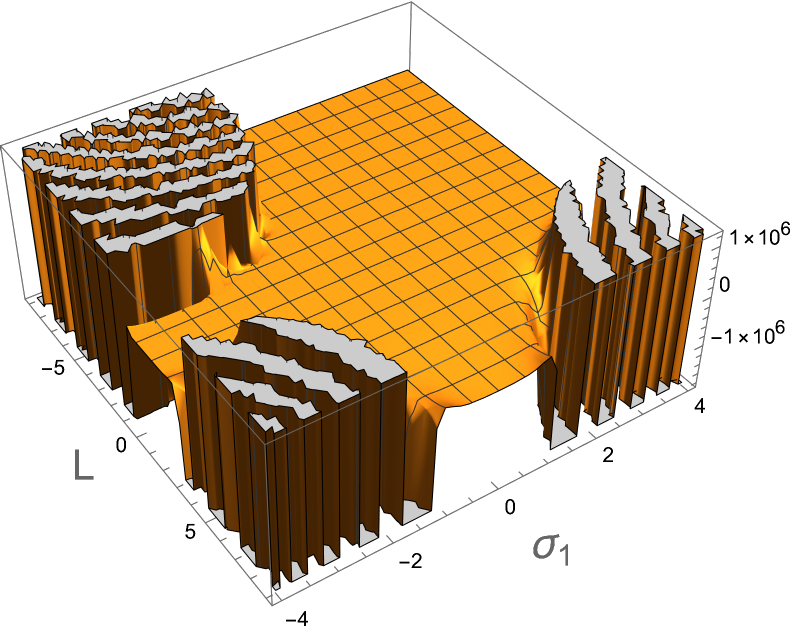}  
\caption{The behavior of $\text{AdS}_2$ partition function integrand versus $L$ and $\sigma_1$ is shown, for the case of $L_0=g_{YM}=\rho_0=r=1$. }
\label{fig:AdS2Partition}
\end{center}
\end{figure}

\begin{figure}[ht!]   
\begin{center}
\includegraphics[width=0.45\textwidth]{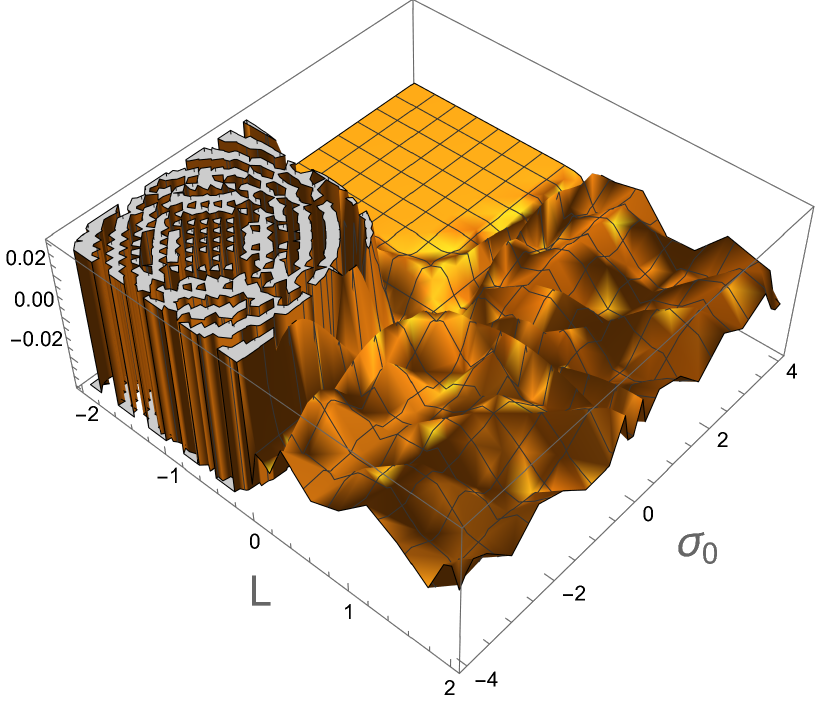}  \ \ \ \ \ \ \ \ \ \ \ 
\includegraphics[width=0.45\textwidth]{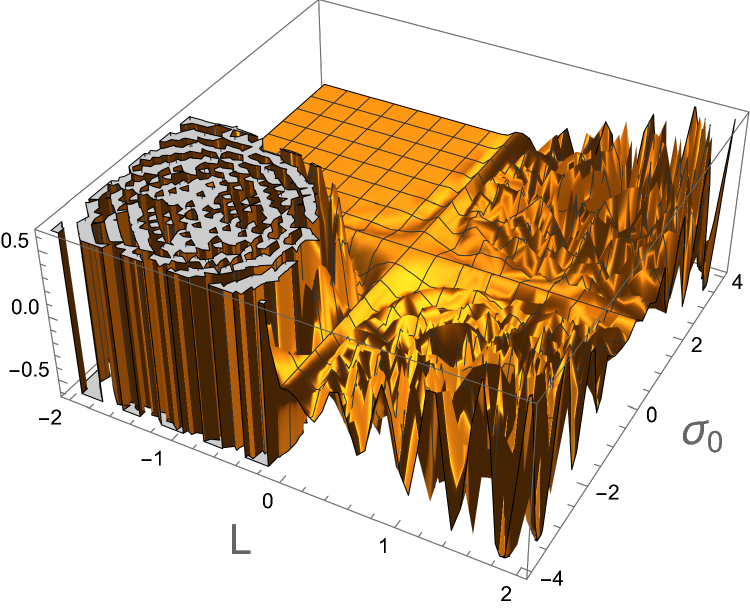}  
\caption{The behavior of $\text{S}^2$ partition function integrand versus $L$ and $\sigma_0$ for the case of $L=\nu=\xi=1$. In the left panel, $r=1$, and in the right panel, $r=2$.}
\label{fig:S2Partition}
\end{center}
\end{figure}

On the other hand, the behavior of the integrand of the partition function found in \cite{GonzalezLezcano:2023cuh} for the case of $\text{AdS}_2$ and $S^2$ has been shown in Figures \ref{fig:AdS2Partition} and \ref{fig:S2Partition} respectively, where for larger values of $\sigma_0$ and $L$, highly oscillatory boundary behavior can be observed in their partition functions. In the case of $\text{AdS}_2$, in Figure \ref{fig:AdS2Partition}, one could see that increasing $r$, $L_0$, and $g_{YM}$ does not change the general behavior of this function. Increasing $\rho_0$ up to 4 preserves the qualitative behavior,  but it is the parameter with largest effect.

In Figure \ref{fig:S2Partition}, for the case of $\text{S}^2$ partition function, one could see that increasing $r$, $L_0$, and $g_{YM}$, does not change the general behavior of this function. Increasing $\rho_0$ up to 4 preserves the qualitative behavior, but it is the parameter with the largest effect.

\begin{figure}[ht!]   
\begin{center}
\includegraphics[width=0.45\textwidth]{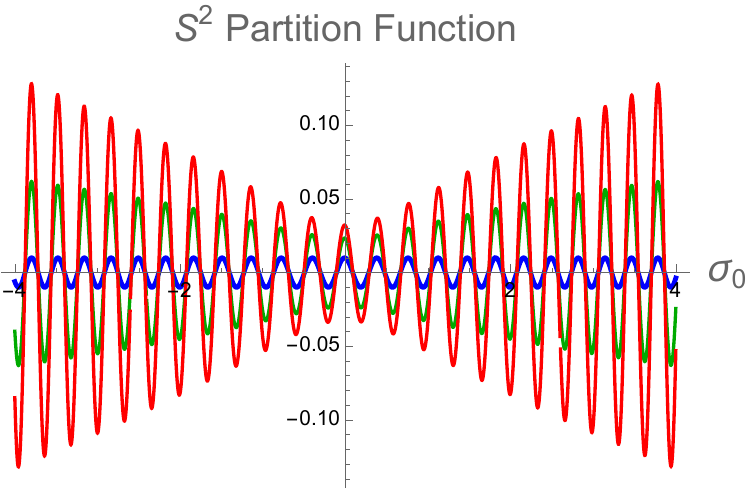}  
\caption{The behavior of $\text{S}^2$ partition function integrand versus $\sigma_0$ for the case of $L=\nu=\xi=1$. The blue curves, whose maxima and minima are constant, correspond to $r=1$, the green curves correspond to $r=1.5$ and the red curves correspond to $r=1.7$. For the case of $r>1$, the absolute values of the maxima and minima would increase with increasing $\sigma_0$.}
\label{fig:S2oscill}
\end{center}
\end{figure}

In Figure \ref{fig:S2oscill},  the oscillatory behavior of $\text{S}^2$ partition function integrand versus $\sigma_0$  is shown. For the case of $r>1$, the absolute values of the maxima and minima would increase with increasing $\sigma_0$.

Interestingly, in \cite{Chen:2023mbc}, for the case of a black hole in $\text{AdS}_4$, similar to our studies here, rapid fluctuations in the quantity $\text{Tr}_E (-1)^F$  as the function of energy $E$ have been detected, where the growth of its envelope has the functional behavior of $\sqrt{E}$. This oscillation is mainly due to the imaginary part of the action, where its frequency is determined by $\text{Im} (I)$, while determining the overall phase would require knowledge of quantum effects and the one-loop determinant.

One could note that increasing the R-charge, $r$, in the $S^2$ partition function, makes the integrand smoother near the center, i.e., $\sigma_0 \to 0$ and $L \to 0$, but enhances the oscillation in the oscillatory regions. This is due to the fact that particle and energy scatter away from the center. In addition, it is worth mentioning here that this oscillatory behavior comes only from the $\Gamma$ functions in the partition function of $S^2$. This is indeed similar to the case of JT, as show in Figure \ref{fig:JT2partPartition}.

\begin{figure}[ht!]   
\begin{center}
\includegraphics[width=0.99\textwidth]{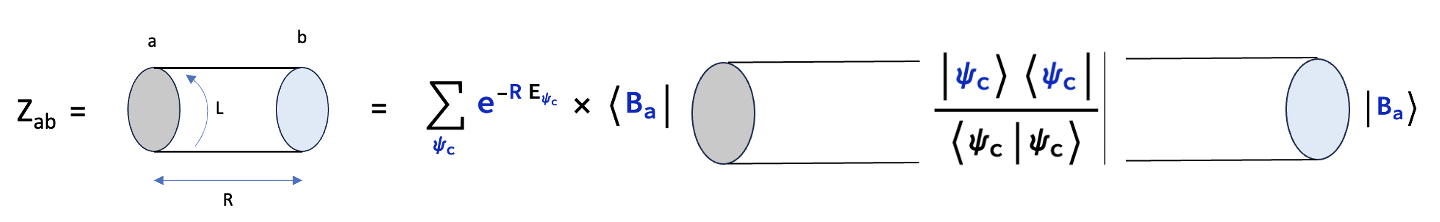}  
\caption{The behavior of the partition function of a cylinder with a circumference $L$ and a length $R$, as described in \cite{Jiang:2019xdz}. The closed string channels which is also related to wavy red line in Fig. \ref{fig:3point} is related to our universal boundary oscillatory behavior and the propagation factor $e^{-R E_{\psi_c}}$ is related to the bulk damping contribution in the partition function.}
\label{fig:stokes12}
\end{center}
\end{figure}

In \cite{Jiang:2019xdz}, the behavior of the partition function of a cylinder with a circumference $L$ and a length $R$ has been studied, where it has been claimed that it accepts two different expansion. In the first case, it could be written as 
\begin{gather}
Z_{ab} = \sum_{\psi_c : \text{closed strings}} e^{-R E_{\psi_c} (L) } \frac{\langle B_a | \psi_c \rangle  \langle \psi_c | B_a \rangle}{ \langle \psi_c | \psi_c \rangle},
\end{gather}
where in the limits of large $R$, the expansion would be $Z_{ab} \overset{R \to \infty}{\sim} e^{-R E_\Omega} g_a g^*_b$, with
\begin{gather}
g_a \equiv \frac{\langle B_a | \Omega \rangle }{\sqrt{\langle \Omega | \Omega \rangle}}, \ \ \ \ g_b \equiv \frac{\langle B_b | \Omega \rangle }{\sqrt{\langle \Omega | \Omega \rangle}}.
\end{gather}

As shown in Fig. \ref{fig:stokes12}, the partition function of cylinder has two channels. One is related to the overlaps between the boundary states and the closed string states, which gives us the oscillatory behaviors, and the one with the propagation factor of $e^{-R E_{\psi_c}}$, which is related to damping brane states.

\begin{figure}[ht!]   
\begin{center}
\includegraphics[width=0.55\textwidth]{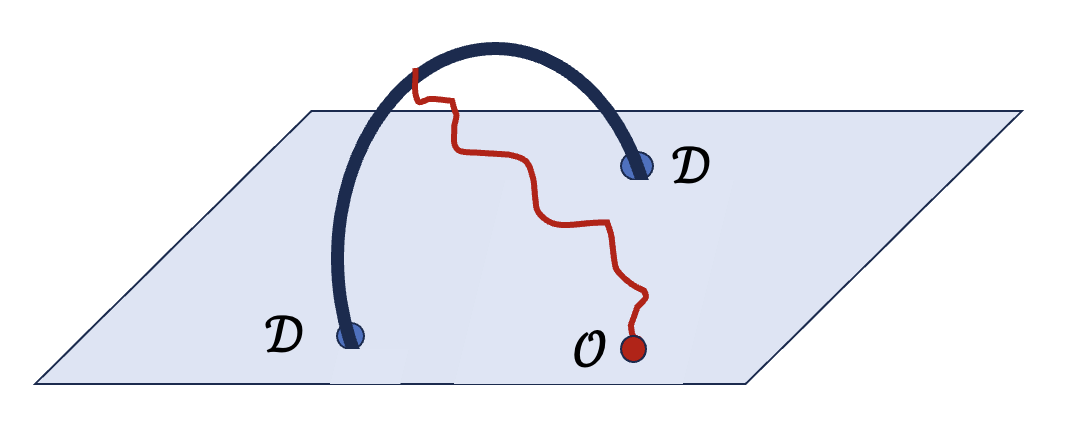}  
\caption{The behavior of AdS three-point function. The red wavy line corresponds to a closed string related to single-trace operator which again is related to our oscillatory behavior.}
\label{fig:3point}
\end{center}
\end{figure}

Note that this partition function is related also to four point-function  which is determined by the sum of exchanges of closed strong states leading to fluctuations and oscillatory behaviors. Also, in the case of $R \gg 1$, it is related to the overlaps between the boundary states and the ground states again connecting to the oscillatory behaviors. Also, we could detect these oscillatory behaviors to intrinsic nature of three-point function and the behavior of closed strings in the setup as well as shown in Fig. \ref{fig:3point}, similar to what discussed in \cite{Jiang:2019xdz}.

So as we explained, in the case of $2d$ Liouville theory, and for the case of cylinder partition function, the general structure of partition function is \cite{Jiang:2019xdz}
\begin{gather}
Z_{\alpha, \beta} ^{\text{cylinder}} = \langle B_\alpha | e^{- H L} | B_\alpha \rangle \simeq \langle B_\alpha | 0 \rangle  \langle 0 | B_\beta \rangle e^{- E_0 L} =g_\alpha g_\beta e^{- E_0 L},
\end{gather}
where the term $g_{\alpha} g_{\beta}$ corresponds to the boundary part, which gives the oscillatory behavior (Gamma functions), and the $e^{- E_0 L}$ term corresponds to the smooth bulk part, which damps very rapidly.  For the special case also, we have 
\begin{gather}
Z_{\alpha \alpha}^{\text{cylinder}} \simeq |g_\alpha|^2 e^{-E_0 L},
\end{gather}
where $g_\alpha \equiv \langle B\ket{0}$ and $g_\beta \equiv \langle 0\ket{\beta}$.

As discussed in \cite{Jiang:2019xdz}, the fluctuation around the saddle point which is also related to our universal oscillatory behavior in the partition function could be written as 
\begin{gather}
\text{fluctuation} = \left ( \text{det} \frac{1}{i} \frac{\delta^2 S}{\delta \eta \delta \rho} \right )^{-1}.
\end{gather}

Again, one should note that the boundary term, which effectively defines the degrees of freedom, can produce the oscillatory behavior, and there are direct connections between the partition functions of various $2d$ gravity models.

Additionally, using the equivariant localization, it has been found that there is a scaling limit where the partition function of chiral gravity PSL(2,$\mathbb{R}$) Chern-Simons theory reduces to JT \cite{Eberhardt:2022wlc}. The level $k$ in chiral gravity, is proportional to $\ell_{\text{AdS}} /G_N$. Moreover, the higher-genus partition function becomes oscillatory, and thus gravity gives a non-smooth contribution to the partition function. Note that the reason that the oscillatory behavior in $k$ and the local dynamical behavior of the island depend on the full topological properties of the background is due to the Haar-invariance and the delocalized quantum information.

\begin{figure}[ht!]   
\begin{center}
\includegraphics[width=0.45\textwidth]{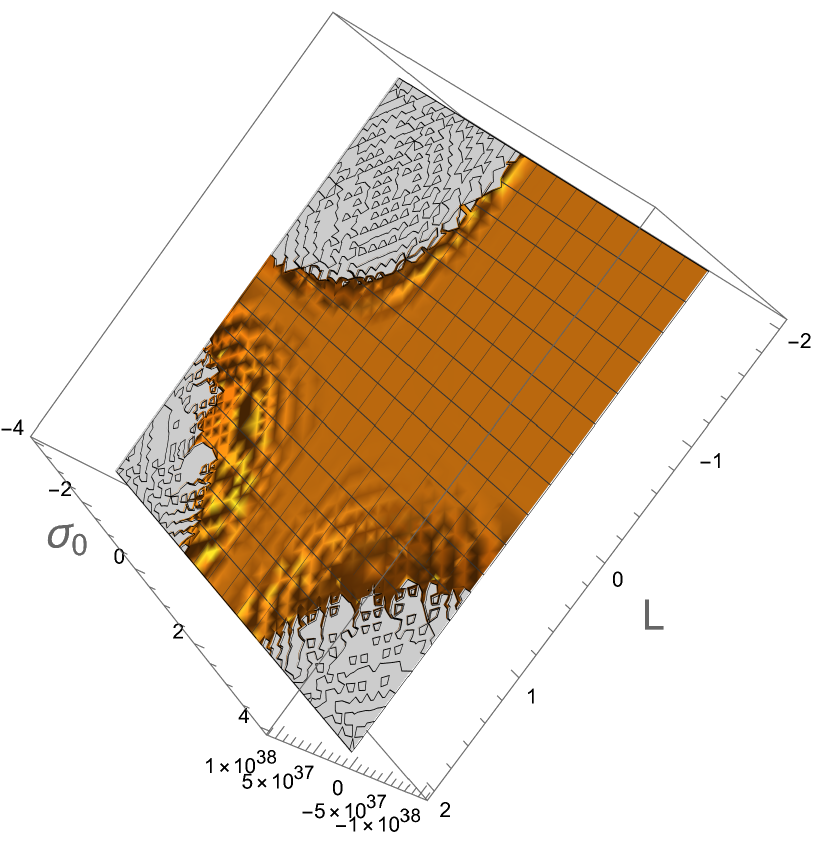}  
\caption{The behavior of the $S^2$ partition function in the large $r$-charge limit.}
\label{fig:bigRcharge}
\end{center}
\end{figure}

For very large values of $r$-charge, as shown in Figure \ref{fig:bigRcharge}, the behavior of the partition function would smooth out. 

\begin{figure}[ht!]   
\begin{center}
\includegraphics[width=0.45\textwidth]{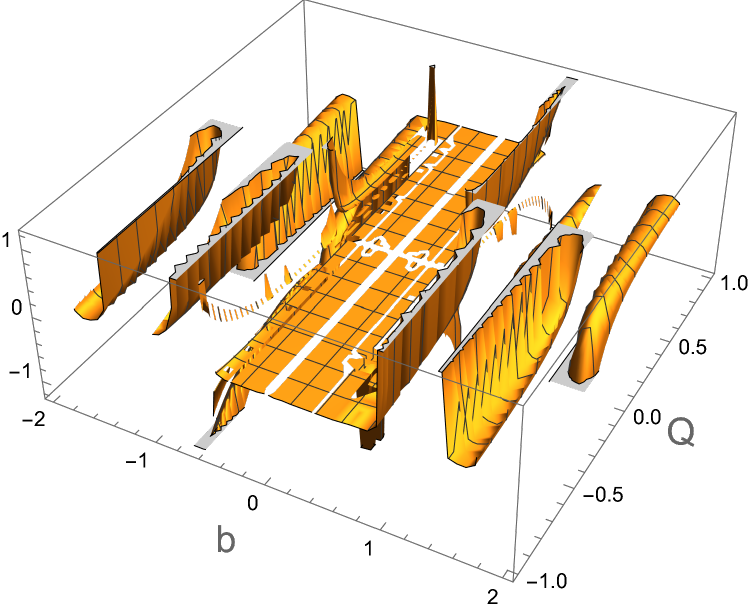}  
\caption{The behavior of the spherical Liouville partition function.}
\label{fig:SpherePF}
\end{center}
\end{figure}

Now we can compare this behavior with other cases. The spacelike spherical partition function for $c_L \ge 25$ is \cite{Giribet:2022cvw}
\begin{gather}
Z\lbrack \Lambda \rbrack = \frac{\left(1-b^2\right) \left(\frac{\pi  \Lambda  \Gamma \left(b^2\right)}{\Gamma \left(1-b^2\right)}\right)^{Q/b}}{\frac{\pi ^3 Q \Gamma \left(b^2\right) \Gamma \left(\frac{1}{b^2}\right)}{\Gamma \left(1-b^2\right) \Gamma \left(1-\frac{1}{b^2}\right)}},
\end{gather}
where its  behavior is shown in Figure \ref{fig:SpherePF}.

\begin{figure}[ht!]   
\begin{center}
\includegraphics[width=0.45\textwidth]{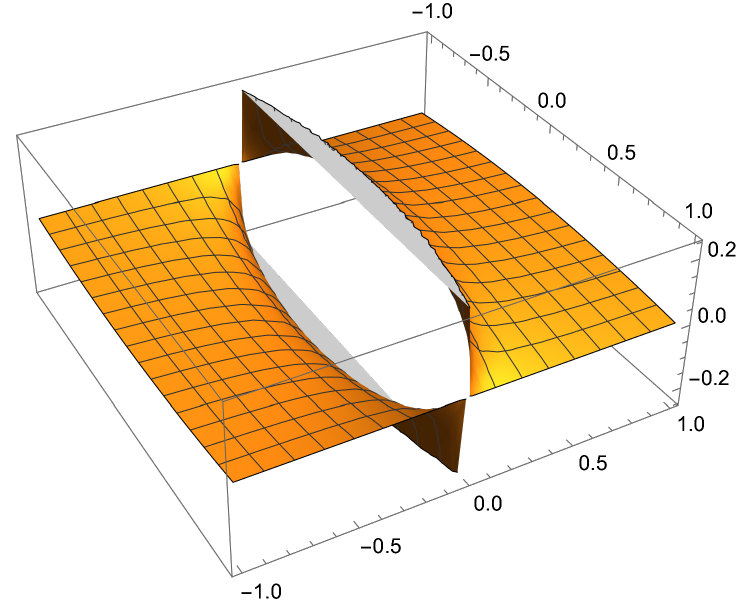}   \ \ \ \ \ \ \ \ \ \ 
\includegraphics[width=0.45\textwidth]{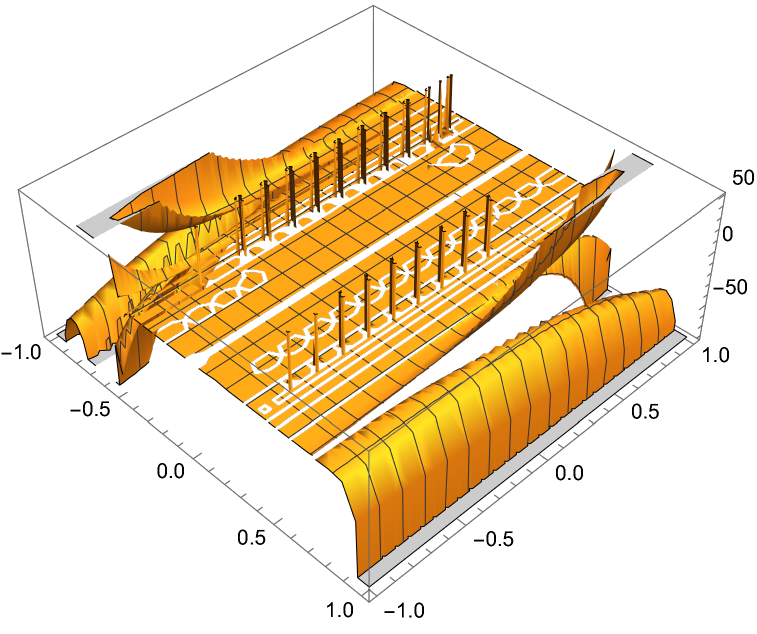}  
\caption{The behavior of the brane state of the spherical Liouville partition function is shown in the left panel and the behavior of its Hartle-Hawking state is shown in the right panel.}
\label{fig:LiouviSph}
\end{center}
\end{figure}

This partition function can again be considered as a combination of the smooth brane states in the bulk, and the boundary oscillatory Hartle-Hawking state, whose behavior is shown in Figure \ref{fig:LiouviSph}.

Also, the timelike Liouville field theory (LFT) partition function on the sphere topology is \cite{Giribet:2022cvw}
\begin{gather}
\hat{Z} \lbrack \Lambda \rbrack = \frac{(1+b^2) (\pi \Lambda \gamma (-b^2 ))^{\hat{Q}/b} }{\pi^3 \hat{Q} \gamma (-b^2) \gamma (-b^{-2}) },
\end{gather}
where this universal behavior which is being able to be written as the combination of Hartle-Hawking states and branes states can be observed in these cases as well.

In addition, in the interesting work of \cite{tHooft:2014znz}, the cellular automaton interpretation of quantum mechanics has been presented, where it is proposed that the states consist of fast-rotating and slow-spectrum components. We suggest here that the fast-rotating case corresponds to the Hartle-Hawking state (adjoint), and the slow one corresponds to the brane state.

\subsection{Eigenfunctions, spectrum and partition functions}

In order to better understand the universalities in low-dimensional gravity partition functions, we can examine the spectrum and quantum Landau levels of a particle moving in $H_2$, with the action \cite{Yang:2018gdb}
\begin{gather}
S= \int du \frac{1}{2} \frac{\dot{x}^2 + \dot{y}^2  }{y^2} +i b \int du \frac{ \dot{x} }{y} - \frac{1}{2} (b^2 + \frac{1}{4} ) \int du, \ \ \ \ \  \ \ \ \ \ \ b=iq.
\end{gather}

If in the above action $q$ is real, it corresponds to an \textit{electric} field, and if $b$ is real, it corresponds to a magnetic field. Then, the Hamiltonian is
\begin{gather}
H= \frac{\dot{x}^2 + \dot{y}^2 }{2y^2}+ \frac{q^2}{2}.
\end{gather} 

Solving the Schrödinger equation for the above Hamiltonian with the condition of vanishing wavefunction at the horizon, $y \to \infty$, and for the case of $k>0$, leads to  \cite{COMTET1987185, Comtet:1984mm, Pioline:2005pf}
\begin{gather}
f_{s,k} (x,y) = \Big(\frac{ s \sinh 2 \pi s}{4 \pi^3 k} \Big)^{\frac{1}{2} } \big | \Gamma ( is - b + \frac{1}{2} ) \big | e^{- i k x} W_{b, is} ( 2 k y), \ \ \  \ \ \ \ \ \ \ \ \omega_{s,k}= \frac{s^2}{2},
\end{gather}
where $W$ is the Whittaker function and $\omega_{ks}$ is the energy of the states labeled by the quantum numbers $s$ and $k$, where $s$ is the quantum number of the continuous series representation of $SL(2)$ with the spin $j=\frac{1}{2} + is$. One can then examine the effects of each parameter on this wave function. 

\begin{figure}[ht!]   
\begin{center}
\includegraphics[width=0.4\textwidth]{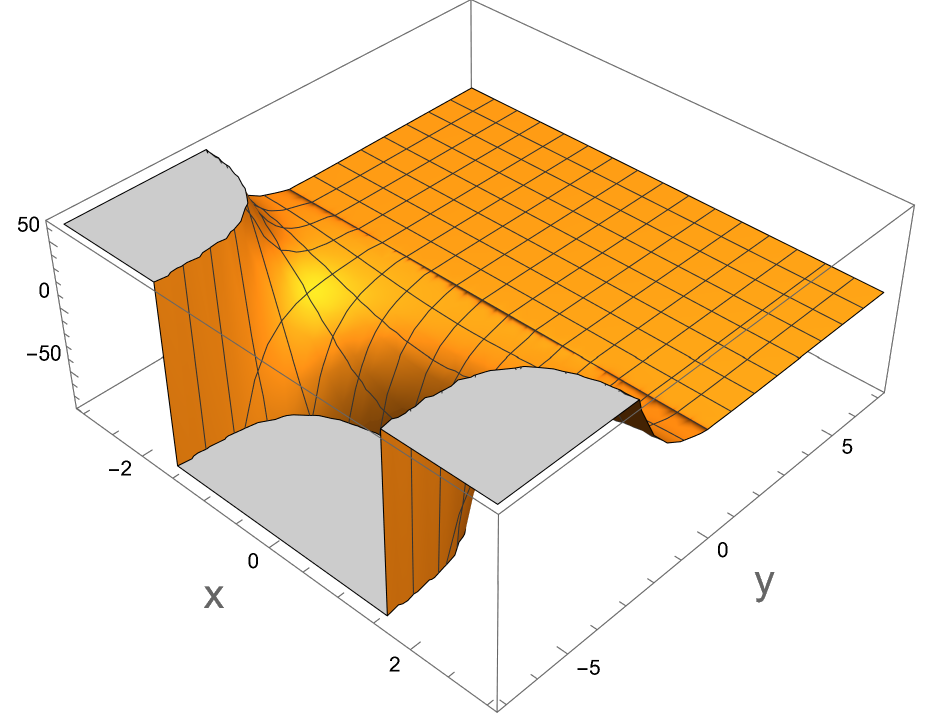} 
\caption{The general behavior of the real part of the wave-function $f_{s,k} (x,y)$ is shown. This is for the case of $s=k=b=1$. }
\label{fig:QMpartitionFucntion}
\end{center}
\end{figure}

\begin{figure}[ht!]   
\begin{center}
\includegraphics[width=0.3\textwidth]{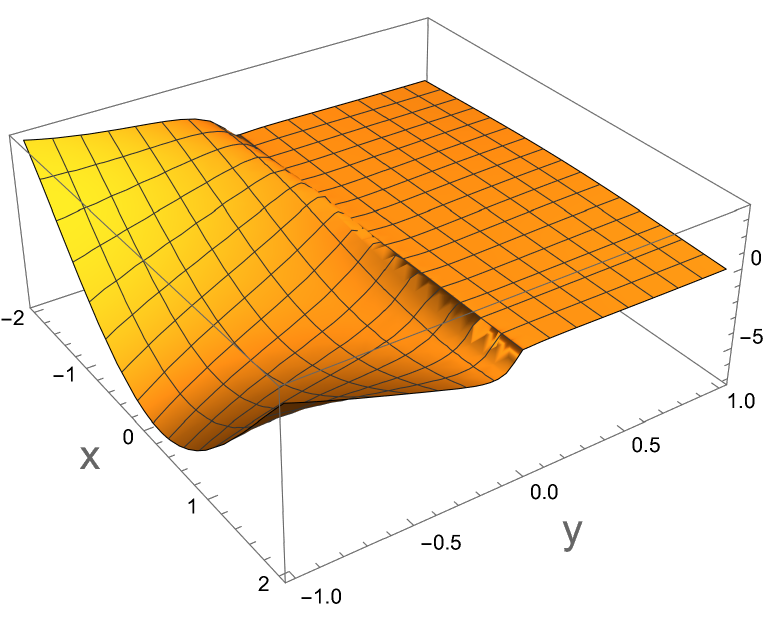} 
\includegraphics[width=0.3\textwidth]{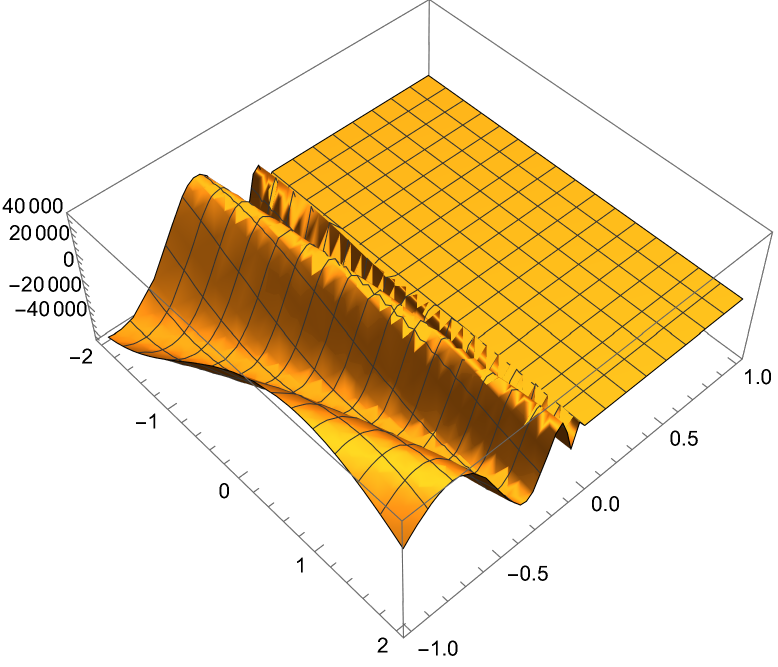}
\includegraphics[width=0.3\textwidth]{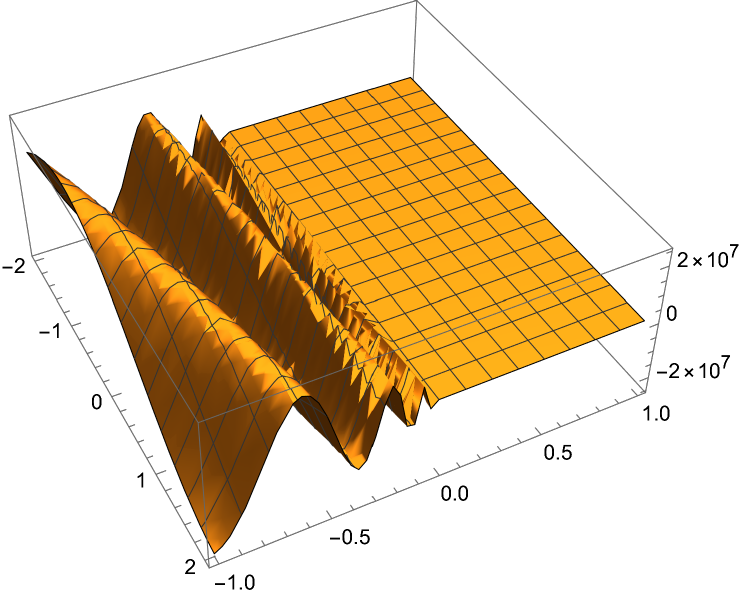}  
\caption{The behavior of the real part of the wavefunction $f_{s,k} (x,y)$ is shown. This is for the case $b=k=1$, while $s=2$ in the left panel, $s=4$ in the center, and $s=6$ in the right section. It can be seen that increasing $s$, would increase the oscillation, as it increases the kinetic energy, while it has no effect in the negative values of $y$. }
\label{fig:fsk}
\end{center}
\end{figure}

\begin{figure}[ht!]   
\begin{center}
\includegraphics[width=0.35\textwidth]{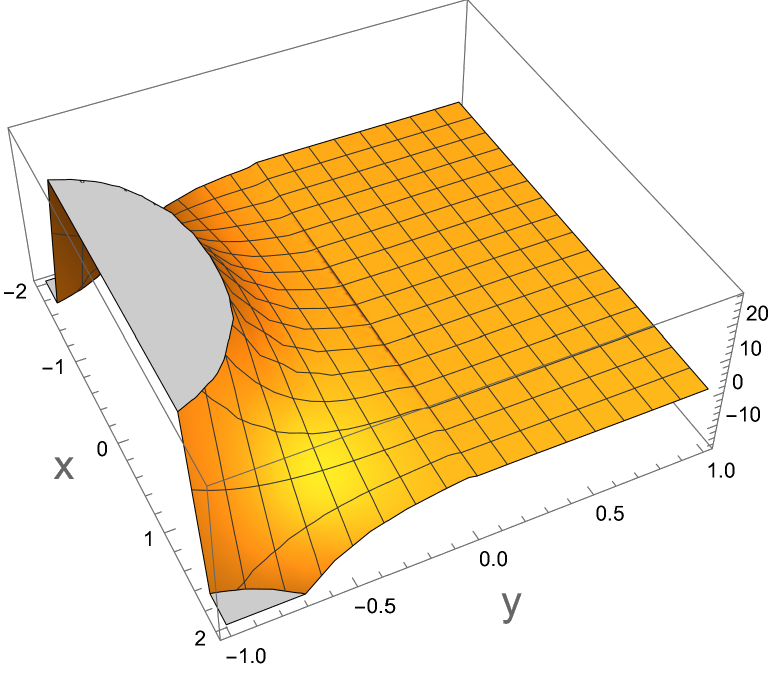} \ \ \ \ \ \ \ \ \ \ \  \ \ \ 
\includegraphics[width=0.35\textwidth]{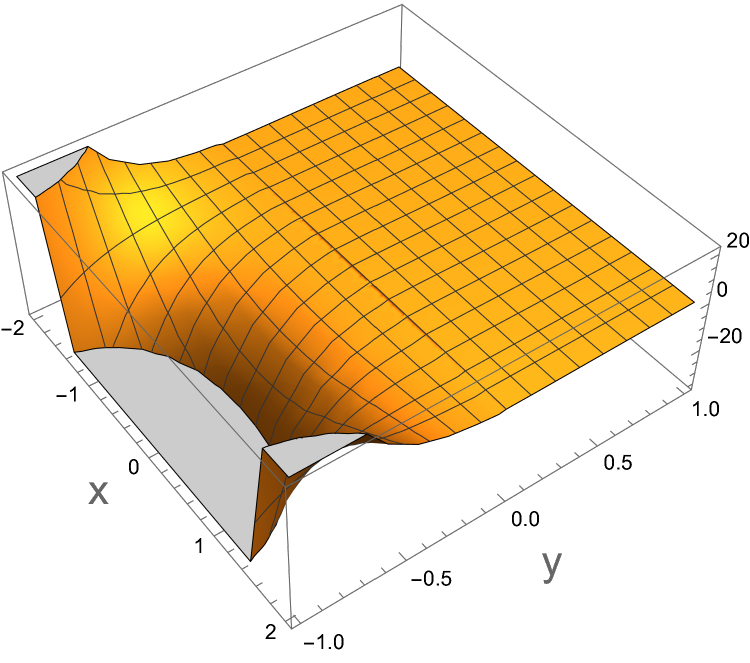}
\caption{The behaviors of the wavefunction $f_{s,k} (x,y)$ for even and odd values of $b$ are shown. On the left, $b=6$, and on the right, $b=7$. The general behavior for other values is similar.  Here, $s=k=1$. }
\label{fig:boddeven}
\end{center}
\end{figure}

From Figure \ref{fig:fsk}, one can see that by increasing $s$, the perturbations in a certain range increase. Also, the wavefunction vanishes at $y \to 0$.
In addition, the effects of the parameter $b$ would depend on the fact that whether it is even or odd. For the odd values of $b$, the partition function attains a minimum, and for even values of $b$, it attains a maximum. When $b$ increases significantly, the dip or peak becomes stretched along the $x$ axis. The behavior of the wavefunction, for even and odd values of $b$ is shown in Figure \ref{fig:boddeven}. 

\begin{figure}[ht!]   
\begin{center}
\includegraphics[width=0.35\textwidth]{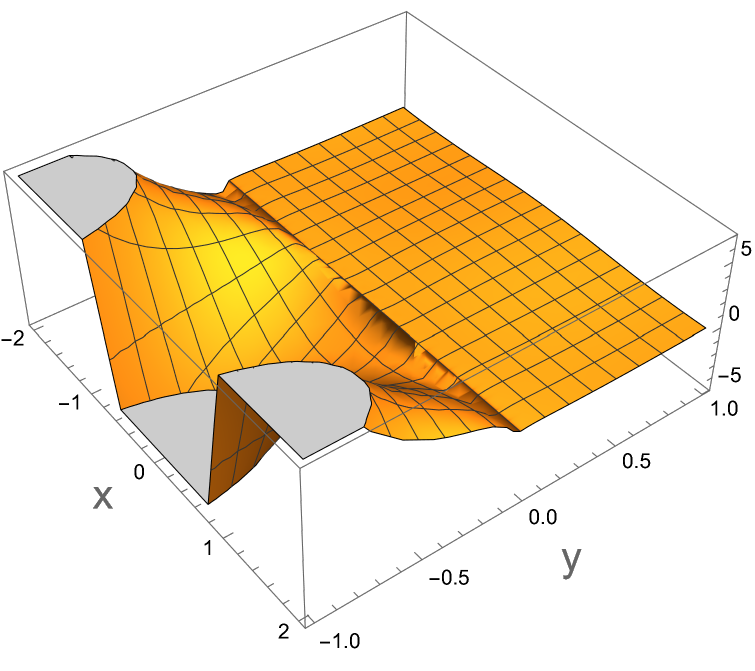}  \ \ \ \ \ \ \ \ \ \ \  \ \ \ 
\includegraphics[width=0.35\textwidth]{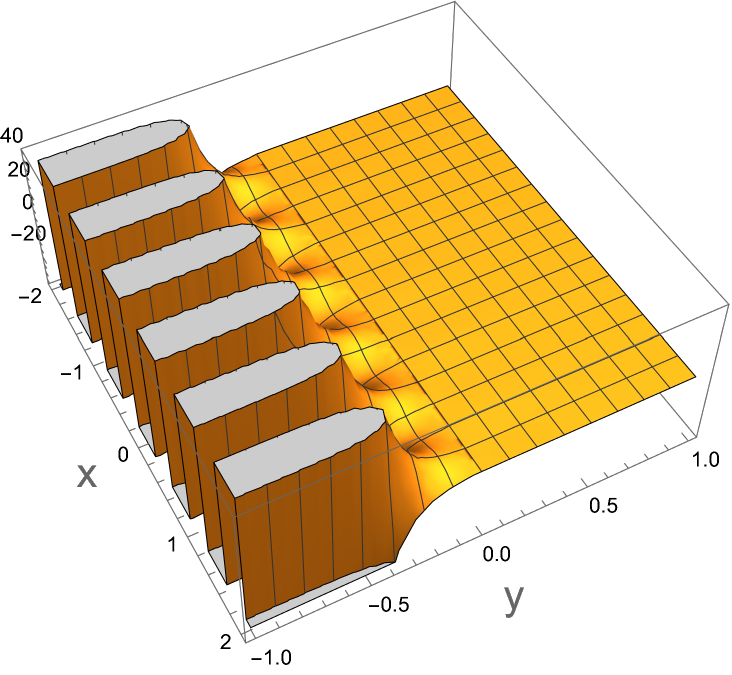}
\caption{The behavior of the wavefunction $f_{s,k} (x,y)$ for different values of $k$ is shown. On the left, $k=2$, and on the right, $k=9$. Here $s=b=1$.}
\label{fig:k}
\end{center}
\end{figure}

The effects of the quantum number $k$ can also be studied here, where the results are shown in Figure \ref{fig:k}. One could see that it is actually the quantum number $k$ that defines the number of peaks in the partition function.

For this quantum mechanical system, the canonical partition function is \cite{Yang:2018gdb}
\begin{gather}
Z_{\text{particle}} = \int_0^\infty ds \int_{0}^\infty dk \int_M \frac{dx dy }{y^2} e^{- \beta \frac{s^2}{2} }f^*_{s,k} (x,y)  f_{s,k} (x,y)\nonumber\\
= V_{AdS} \int_0^\infty ds e^{- \beta \frac{s^2}{2} } \frac{s}{2\pi} \frac{\sinh (2\pi s) }{\cosh(2\pi q)+ \cosh(2\pi s) },
\end{gather}
which again has the same structure as we have seen before and as shown in Figure \ref{fig:QMpartitionFucntion}, with a smooth exponential function and a non-smooth oscillatory function consisting of $\sinh (x)$ and $\cosh(x)$,

\begin{figure}[ht!]   
\begin{center}
\includegraphics[width=0.31\textwidth]{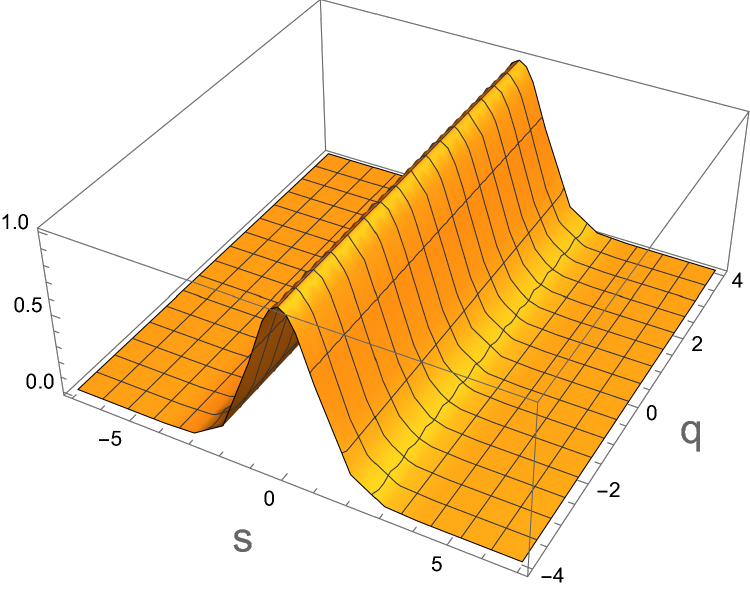} 
\includegraphics[width=0.31\textwidth]{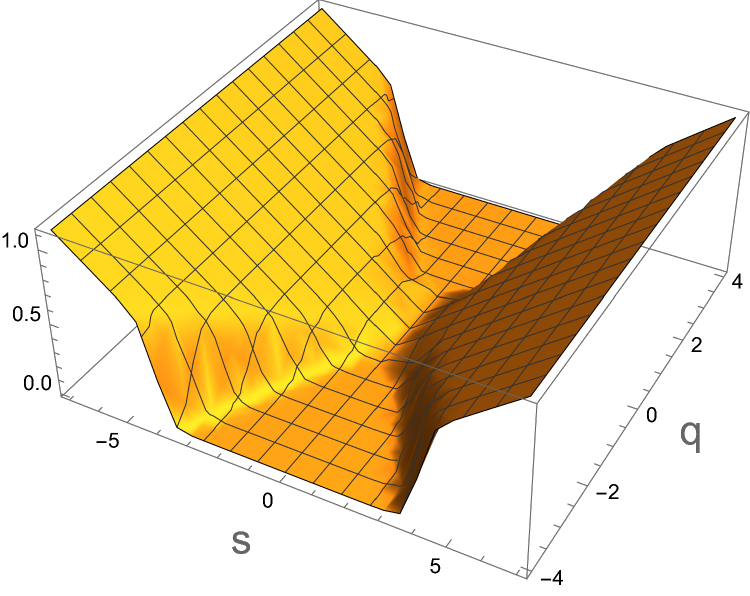}   ${\to}$
\includegraphics[width=0.31\textwidth]{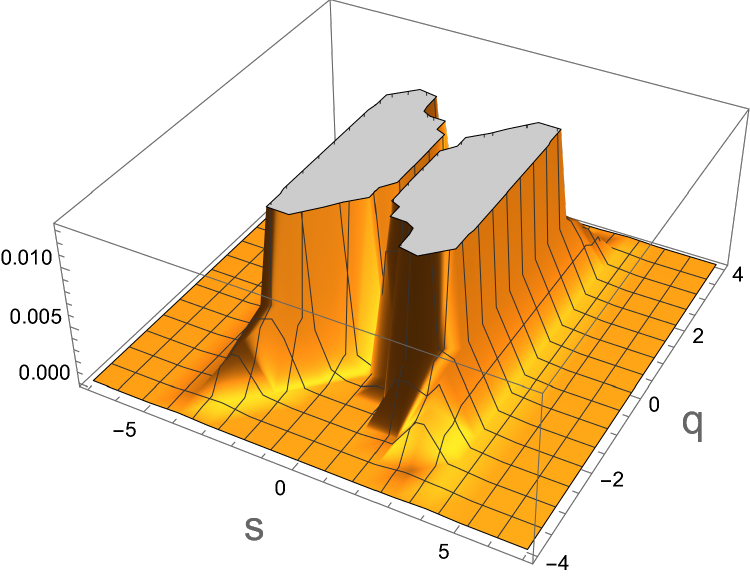}
\caption{On the left, the exponential part; in the middle, the oscillatory and non-smooth part; and on the right, the full behavior of the action is shown.  The universal behavior can be noted as expected.}
\label{fig:QMpartitionFucntion}
\end{center}
\end{figure}

Now, we can interpret the results of the partition function of $S^2$ and $\text{AdS}_2$ using the picture of a non-relativistic particle in an electric or magnetic field as well. The most important point that needs to be clarified in these examples is the meaning of the quantum number $k$, which satisfies $H \ket{j,k} = j(1-j) \ket{j,k}$ and $L_{-1} \ket{j,k} = k \ket{j,k}$. In the case of supersymmetric models, this is related the $R$-symmetry group and $r$-charge.

In this case, the magnetic field $q$ in the JT case is related to the R-symmetry flux or FI parameter $\xi$ in the $\mathcal{N}=(2,2)$ case. The energy $s^2/2$ is related to the Coulomb branch parameter $\sigma_0 L$ or $\sigma_1 L$. The particle mass $\mu$ is related to the R-charge $r$ or $\rho_0 L$, and the boundary length $\ell$ is related to the background size $L/L_0$. Also, in the case of measure, we should have the following relation
\begin{gather}
\int \frac{d(\sigma_0 L)}{(g_{YM} L_0 )^2} \ \ \text{or} \ \ \int \frac{d(\sigma_1 L) }{g_{YM} L_0 }  \ \to \ \int_0^\infty ds \rho(s)=\int_0^\infty ds \frac{s}{2\pi^2} \sinh (2\pi s).
\end{gather}

\subsection{Wheeler-DeWitt wavefunctions and universalities}

In the previous sections, the Hartle-Hawking state and Wheeler-DeWitt wavefunctions have been used to extract and analyze the universalities observed in the partition function and in the behavior of low-dimensional quantum gravity models.

In the recent work of \cite{Chua:2023ios}, the Hartle-Hawking state and its factorization in $3d$ has also been studied, where the connections to Alekseev-Shatashvili theories and Liouville theory have been used. In Liouville theory, specifically, the Hartle-Hawking state and Wheeler-DeWitt wavefunctions have been obtained. Similar to \cite{Ghodrati:2022hbb},  the interconnections between $3d$ and $2d$ Liouville theory and JT gravity model have been explained. 

One important observation there is that two copies of Liouville theory in $3d$ gravity behave similarly to the Schwarzian theory in JT gravity.  The similarities between Liouville theory and the Schwarzian theory in JT model have been further studied in \cite{Maldacena:2016upp}. The main point is that the (mixed or entangled) quantum state in two copies of Liouville theory in $2d$ is dual to the Hartle-Hawking state in $3d$ gravity. This signals that there are many more universalities between low-dimensional gravity models (between $2d$ and $3d$ gravity models) than noted previously. As mentioned in \cite{Chua:2023ios}, specifically the Liouville theory can be used as an intermediate model to derive such dualities. This can also be seen from Figure \ref{fig:dimHoltheories} here.

As demonstrated in \cite{Chua:2023ios}, the Hartle-Hawking (HH) state in $3d$ is dual to two copies of the Liouville ZZ boundary state, which can be written in the form of
\begin{gather}
\ket{\Psi_{\beta/2}^{HH}} \cong e^{-\beta H/4} \ket{ZZ} e^{-\beta H /4} \ket{\widetilde{ZZ}},
\end{gather}
which one can also infer from Figure \ref{fig:dimHoltheories}.  Similarly, one would expect that the HH state in $3d$ Chern-Simons theory is dual to two copies of $2d$ WZW theory. Note that $H$ here is the Hamiltonian of the Liouville theory.

The partition function of Liouville theory could also be considered as the combination of two matrix models of the form
\begin{gather}
\tau_N = \mathcal{Z} (N) = \int dM d\bar{M} e^{-N \left(\mathrm{tr} (M \bar{M} ) + \sum_{k>0} (t_k \mathrm{tr} M^k + \bar{t}_k \mathrm{tr} \bar{M}^k ) \right) },
\end{gather}
where the $t_k$s, and $\bar{t}_k$ are Toda times. The same spectral universality behaviors, such as Airy edge behavior and collective eigenvalue description, are present here.

The wave functions are \cite{Betzios:2020nry}
\begin{gather}
\psi^+ (\omega, \lambda) =  \left (\frac{1}{4\pi \sqrt{(1+ e^{2\pi \omega})} } \right)^{\frac{1}{2}} 2^{1/4} \left |\frac{\Gamma(1/4+ i \omega/2)}{\Gamma(3/4+ i \omega/2)} \right |^{1/2} e^{-i \lambda^2/4} {}_1F_1(1/4-i \omega/2,1/2;i\lambda^2/2) \nonumber\\
=\frac{e^{-i \pi/8}}{2\pi} e^{- \omega \pi/4} | \Gamma (1/4 + i \omega/2) | \frac{1}{\sqrt{|\lambda|}} M_{i\omega/2, -1/4} (i \lambda^2/2), \nonumber\\
\psi^- (\omega, \lambda) =  \left (\frac{1}{4\pi \sqrt{(1+ e^{2\pi \omega})} } \right)^{\frac{1}{2}} 2^{3/4} \left |\frac{\Gamma(3/4+ i \omega/2)}{\Gamma(1/4+ i \omega/2)} \right |^{1/2}\lambda  e^{-i \lambda^2/4} {}_1F_1(3/4-i \omega/2,3/2; i\lambda^2/2) \nonumber\\
=\frac{e^{ -3 i\pi/8}}{\pi} e^{- \omega \pi/4} | \Gamma (3/4 + i \omega/2) | \frac{\lambda}{|\lambda|^{3/2}} M_{i\omega/2, 1/4} (i \lambda^2/2),
\end{gather}
where, similar to Wheeler-DeWitt wavefunctions in JT gravity, they have oscillatory WKB region and an exponential barrier region, and in both cases the solutions are modified Bessel-type functions. Specifically, in Liouville minisuperspace, one gets
\begin{gather}
\Psi_P (\phi) \sim K_{2iP/b} ( 2\sqrt{\mu} e^{b\phi}),
\end{gather}
so both cases belong to the same function universality class, especially near the boundary.

Also, in JT gravity,  the Schwarzian density of states corresponds to the reflection part, while similarly Liouville and other low-dimensional models have a reflection amplitude that encodes the boundary scattering data. Moreover, the Wheeler-DeWitt wavefunction in both cases contains baby-universe and genus-expansion effects through matrix model connections,  so in all these cases they include disconnected geometries and wormholes. In another recent paper \cite{Hartnoll:2022snh}, Hartnoll used Wheeler-DeWitt states of the AdS-Schwarzschild interior to probe inside the black hole. In fact, the universalities found in the Wheeler-DeWitt states and partition function could help determine the behavior of states inside black holes as well.

Another important observation is that one could expect the dimensional reduction along the angular $U(1)$ or $SO(2)$ orbits, with truncation to circle-invariant fields could change $3d$ Hartle-Hawking states onto a $2d$ slice, which then it would lead to states in JT theory. In the gauged formalism, the projection of states in $3d$ Chern-Simons theory would similarly lead to $2d$ BF theory. These relations among higher dimensional theories and lower ones are the roots of the universalities we found here, especially between the partition functions of these low-dimensional theories. All of these points and universalities can also be seen from the behaviors of the Wheeler-DeWitt wavefunctions in these gravity models . In \cite{Harlow:2018tqv}, the authors discuss the corresponding reduction of the Hartle-Hawking/Wheeler-DeWitt state for specific boundary data. Likewise, the reduction
\begin{gather}
A_3 = A_2+ X d\varphi, \ \ \ \ \partial_\varphi A_3=0,
\end{gather}
takes the Chern-Simons action to a two-dimensional BF action, up to boundary terms and global sectors. However, one should note that the corresponding map between quantum states is not automatic, as the retained sector, boundary conditions, holonomies, etc, must be specified. Therefore, this restricted dimensional-reduction result cannot by itself be taken as the origin of all the broader partition function universality we notice here and more rigorous studies are needed.

Here, we can use the results of \cite{Yang:2018gdb}, where the main point is the calculation of all-order correlation functions in $2d$ gravity, specifically in JT gravity. It can be extended to other $2d$ gravity backgrounds, such as supersymmetric models, and even $3d$ cases, and therefore, further universalities can be found in this way. The general forms and universal behaviors across these cases can then be examined.

The main formula found in \cite{Yang:2018gdb} is
\begin{gather}
\langle O_1 (u_1) ... O_n(u_n) \rangle_{\text{QG}} = \int_{x_1 >x_2...>x_n}  \frac{ \prod_{i=1}^{n} dx_i dz_i  }{V(\text{SL}(2,\text{R}))} \tilde{K} (u_{12}, \boldsymbol{x}_1,\boldsymbol{x}_2)  ...  \tilde{K} (u_{n1}, \boldsymbol{x}_n,\boldsymbol{x}_1) \times \nonumber\\ z_1^{\Delta_1 -2} . . z_n^{\Delta_n -2} \langle O_1 (x_1) ... O_n (x_n) \rangle_{\text{CFT}}. 
\end{gather}

In the above relation, the correlation function in $2d$ JT gravity has been written in terms of the correlation functions of the QFT in hyperbolic space or $\text{AdS}_2$. One could also consider writing the same relation for the Liouville or BF models. In the above relation, the factor $\frac{1}{\text{V}( \text{SL} (2,\text{R}))}$ specifies that the $\text{SL}(2,R)$ gauge symmetry should be fixed, which for other theories and other scenarios implies that the specific symmetry in that case should then be considered.

One important point about JT gravity being simulated by relativistic charged particle is its behavior in the limit of large charges, i.e., $q \to \infty$. In such limits, the propagator acquires a finite part multiplied by a step function $\theta(x_1-x_2)$. One could find the same limits for other $2d$ or $3d$ gravity models. 
In principle, the structure of the propagator $\tilde{K} (u, \boldsymbol{x} _1, \boldsymbol{x}_2)$  would be of the form of $e^{-2 \frac{z_1+z_2}{x_1=x_2} } f(u, \frac{z_1 z_2}{(x_1 -x_2)^2} )$, which for the case of JT is a product of the $SL(2)$ symmetry.  This general form is universal in other theories as well.  

These universalities can also be found in the behavior of propagators.  For the case of JT, which is equivalent to a relativistic charged particle in an electric or magnetic field, the propagator $K(u,\boldsymbol{x} _1, \boldsymbol{x}_2) = \langle \boldsymbol{x} _1 | e^{-u H} | \boldsymbol{x}_2 \rangle $ has been found in \cite{COMTET1987185, Yang:2018gdb}, and can be written as
\begin{flalign}
G(u, \boldsymbol{x} _1, \boldsymbol{x}_2) &= e^{ i \varphi (\boldsymbol{x} _1, \boldsymbol{x}_2)} \int_0^\infty ds s e^{-u \frac{s^2}{2} } \frac{\sinh 2 \pi s}{2\pi (\cosh 2 \pi s + \cos 2 \pi b) } \frac{1}{d^{1+ 2 i s} }\times \nonumber\\ &
\times \  {}_2F_1( \frac{1}{2} - b + i s, \frac{1}{2}+ b + is, 1, 1- \frac{1}{d^2} ).\nonumber\\
d &= \sqrt{\frac{(x_1 - x_2)^2 + (y_1 + y_2)^2}{4 y_1 y_2} },\nonumber\\
e^{i \varphi (\boldsymbol{x} _1, \boldsymbol{x}_2) } &= e^{- 2 i b \arctan \frac{x_1 -x_2}{y_1 + y_2} }.  
\end{flalign}

The analogue of Wheeler-DeWitt wavefunction in the Schwarzian limit for other $2d$ and $3d$ gravity models such as BF, Liouville, and Chern-Simons can also be analyzed in the framework of \cite{Harlow:2018tqv}. Using the behavior of Wheeler-DeWitt wavefunctions, the universal linear growth of complexity and correspondingly the ERB bridge in quantum gravity systems can be shown, which we further study in the next section.

\section{RG flows, entanglement and partition functions: Universalities in wormholes} \label{sec:RGflowsEE}

In this section, we use entanglement entropy, complexity and mixed correlation measures to construct renormalization group flow across dimensions, between field theories with different couplings, and across distinct sectors of a given theory. The goal is to further identify universalities in the structures of quantum gravity models and their wormhole saddles. In particular, we draw on the example of wormholes in Type IIB string theory on $\text{AdS}_5 \times \text{S}^5$ studied in \cite{Cotler:2021cqa, Cotler:2022rud}.

The flows across dimensions have been recently constructed in \cite{GonzalezLezcano:2022mcd}, and the idea of coupling two field theories in different sectors such as $SU(N)$ to a field theory with $SU(N+1)$ or to $SU(M)$ has been studied in \cite{Cotler:2022rud} and \cite{Cotler:2021cqa}, where the authors connected two theories by averaging their partition functions. The joint energy constraints found in \cite{Cotler:2022rud} are written in the form of delta functions in the path integral formulation, which lead to new saddles. These saddles are actually connected to the null energy conditions used in \cite{GonzalezLezcano:2022mcd} to produce the new RG flows.

Actually, different kinds of wormhole solutions have been found in the literature; especially the works of \cite{Cotler:2021cqa, Cotler:2022rud} are two main examples. In \cite{Harvey:2023oom}, also the three-dimensional eternal traversable wormhole deformations of BTZ black hole have been constructed.

\subsection{Entanglement entropy in wormholes}
Here, we examine the RT surfaces, minimal wedge cross sections and the corresponding entanglement entropy and mixed correlation measures in the background of wormholes and black holes, in order to probe various saddles, phase transitions and universalities in the wormhole structures.

In the case of one-boundary AdS soliton, we have the following geometry
\begin{gather}\label{eq:worm1}
ds^2= d\rho^2 + \left ( \frac{4\pi}{d L} \right )^2 \cosh^{\frac{4}{d} } \left ( \frac{d \rho}{2} \right) \left(\tanh^2 \left(\frac{d \rho}{2}\right ) (dx^2)^2 + \beta^2 d \tau^2 + \sum_{i=3}^d  (dx^i)^2 \right).
\end{gather}

In this metric, the spatial torus has $x^2 \sim x^2 + L$ relation, and the spatial circle $x^2$ contracts. The normalized action is
\begin{gather}
S_{\text{grav}} = - \frac{V \beta}{4dGL} \left ( \frac{4\pi}{dL} \right)^{d-1}.
\end{gather}

This metric of AdS soliton would be dominate at low temperature with a gap to the black hole threshold.

The kind of new saddle that \cite{Cotler:2021cqa, Cotler:2022rud} have found in this example with two asymptotically AdS boundaries can describe the random matrix theory energy level repulsion of black hole microstates. 

We also have solutions that describe connected two-boundary Type-IIB wormhole background which could be written as
\begin{gather}\label{eq:wormg2}
ds^2= d \rho^2 + \frac{b^2 \cosh(2\rho)-1}{2}  \left(   \left(\frac{ \beta_1 e ^{2 \rho} + \beta_2 e^{-2 \rho}  }{2(\cosh (2\rho) - \frac{1}{b^2} ) } \right)^2 d\tau^2 + d \Omega_3^2 \right ) + d \Omega_5^2,\nonumber\\
 \text{with the fields:} \ \ \ \ e^\Phi=g_s, \ \ F_5=4( -i \text{vol}_5 + \text{vol}_{\mathbb{S}^5}),
 \label{eq:wormg2}
\end{gather} 
where $ \rho_\Lambda <  \rho < \infty$ and $b>1$. 

Here $\beta_1$ and $\beta_2$ are the sizes of the Euclidean time circles on the two boundaries. The parameter $b$ controls the size of the bottleneck of the wormholes and is related to the boundary energy, where the action of the wormholes would depend on this parameter, i.e, $E= b^4 E_0 = \frac{3\pi b^4}{32G}$. Also, $E_0$ is the energy of the lightest non-rotating black hole which has a $S^1 \times S^3$ boundary topology in Euclidean signature.

Note that relations \ref{eq:worm1} and \ref{eq:wormg2} actually have different topologies and cannot be related by a coordinate transformation as discussed in \cite{Harvey:2023oom}. Also, Eq. \ref{eq:wormg2} is a constrained instanton, not an ordinary solution of the unconstrained Type-IIB equations.

We can take the metric in the general form of
\begin{gather}
ds^2=\alpha(\rho)\left \lbrack \beta(\rho) d\rho^2 + dx^\mu dx_\mu \right \rbrack + g_{ij} d\theta^i d\theta^j,
\end{gather}
where $x^\mu (\mu=0,1, ... , d)$ parametrize $\mathbb{R}^{d+1}$, and $\rho$ is the holographic radial coordinate with the domain of $\rho_\Lambda < \rho < \infty$.

However, one should note that the KKM formalism of \cite{Klebanov:2007ws, Kol:2014nqa} is for a strip in a background with $\mathbb{R}^{d+1}$ factor. On the other hand, the wormhole metric \ref{eq:wormg2} has boundary topology $S^1 \times S^3$, and since $\tau$ is the only coordinate in $g_{\mu\nu}dx^\mu dx^\nu$, then $d=0$ leading to a factor where $V_{-1}$. Therefore, we should look for other wormhole geometries to try to use this setup for.

We can try to use this setup for the following metric \cite{Cotler:2020lxj}
\begin{gather}\label{eq:type2worm}
ds^2 = d\rho^2 + \frac{L^2 b^4}{8} \frac{(\beta_1 e^{2\rho/L} + \beta_2 e^{-2\rho/L})^2}{b^2 \cosh(2\rho/L)-1}d\tau^2 + \frac{L^2}{2} \left ( b^2 \cosh \left ( \frac{2\rho}{L} \right ) -1 \right) d\Omega_3^2,
\end{gather}
which is coming from Type IIB supergravity and has the topology of $S_{\beta_1}^1 \times S^3$ in the boundary.  Here we have $\rho_\Lambda= \frac{1}{2} \text{ArcCosh}(1/b^2) \approx 1.547$, so the plot starts just above this point.

If we apply this setup for the wormhole metric we get
\begin{gather}
\alpha( \rho) = \frac{b^2 \cosh(2\rho)-1}{2} \left ( \frac{\beta_1 e^{2\rho}+ \beta_2 e^{-2 \rho} }{ 2 \left ( \cosh(2\rho)- \frac{1}{b^2} \right) } \right)^2,\ \ \ \ \ \ 
\beta( \rho) = \frac{2}{ b^2 \cosh(2\rho)-1} \left( \frac{ 2 \left(\cosh(2\rho) -\frac{1}{b^2} \right) }{\beta_1 e^{2\rho} + \beta_2 e^{-2\rho}  }  \right)^2.
\end{gather}

\begin{figure}[ht!]   
\begin{center}
\includegraphics[width=0.45\textwidth]{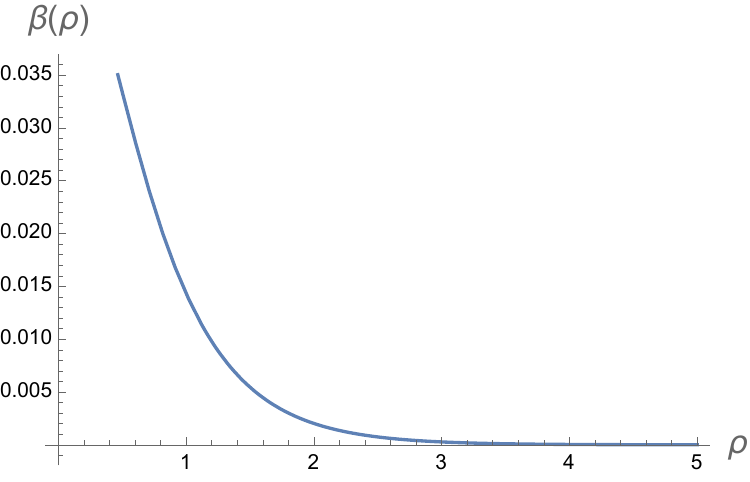} 
\includegraphics[width=0.45\textwidth]{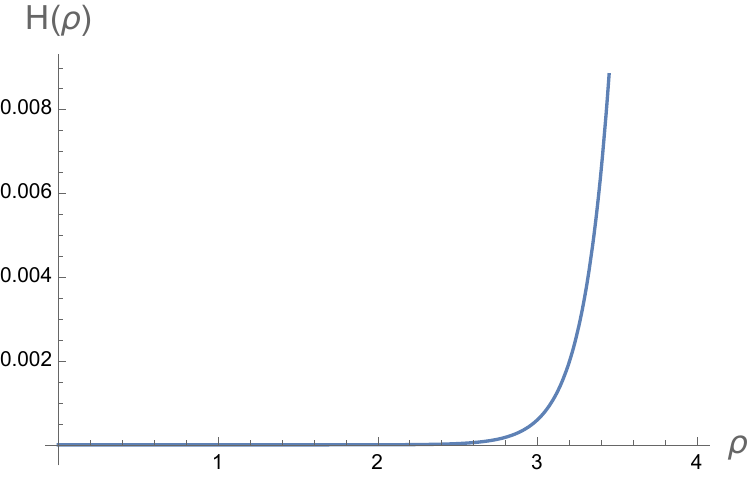} 
\caption{The behavior of the function $\beta({\rho})$ for the wormhole metric, for $b=3$, $\beta_1=\beta_2=2$ is shown in the left panel. If the value of $\beta_1 =\beta_2= n \in \mathbb{R}$, the function $\beta(\rho)$ is always monotonically decreasing. The behavior of the function $H({\rho})$ for the wormhole metric, for $b=3$ is shown in the right section.}
\label{fig:betawormhole3}
\end{center}
\end{figure}

\begin{figure}[ht!]   
\begin{center}
\includegraphics[width=0.4\textwidth]{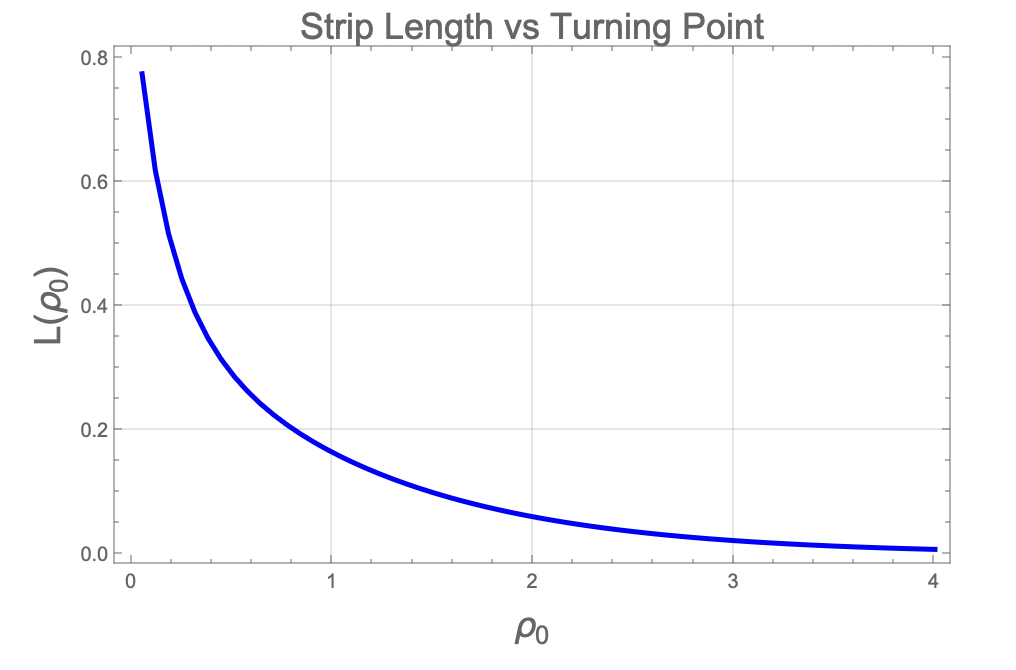}  \ \
\includegraphics[width=0.43\textwidth]{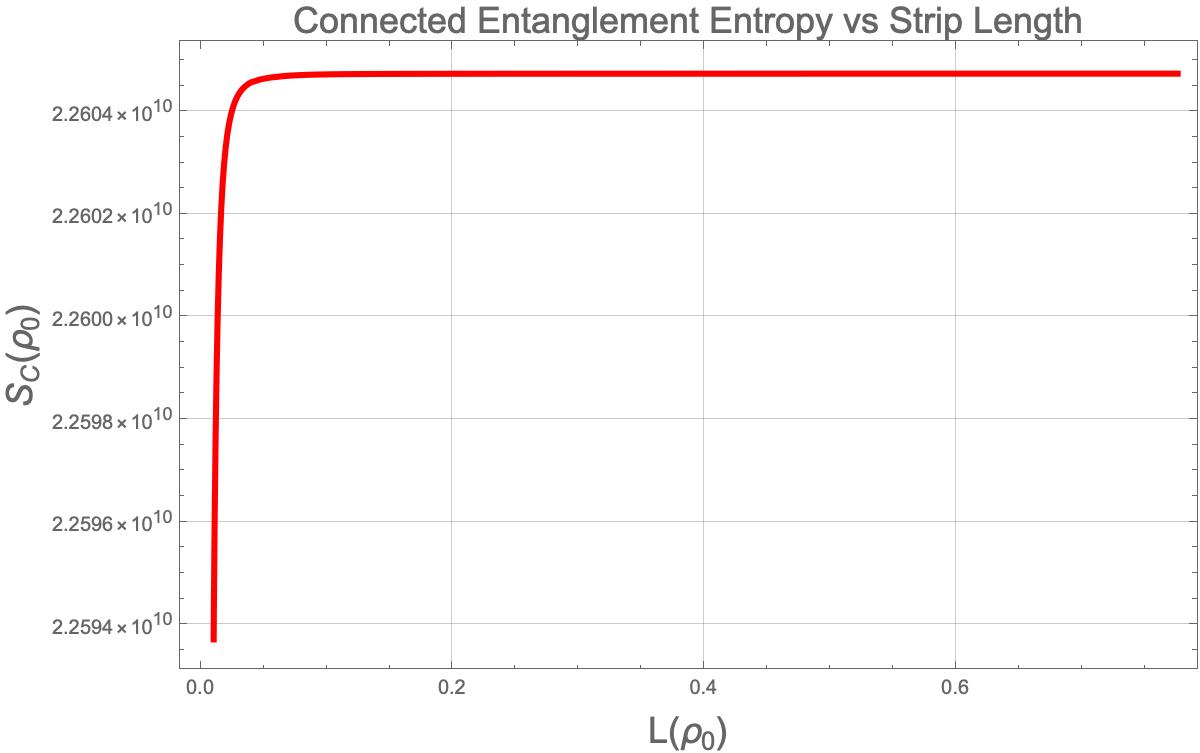}  \ \ 
\includegraphics[width=0.4\textwidth]{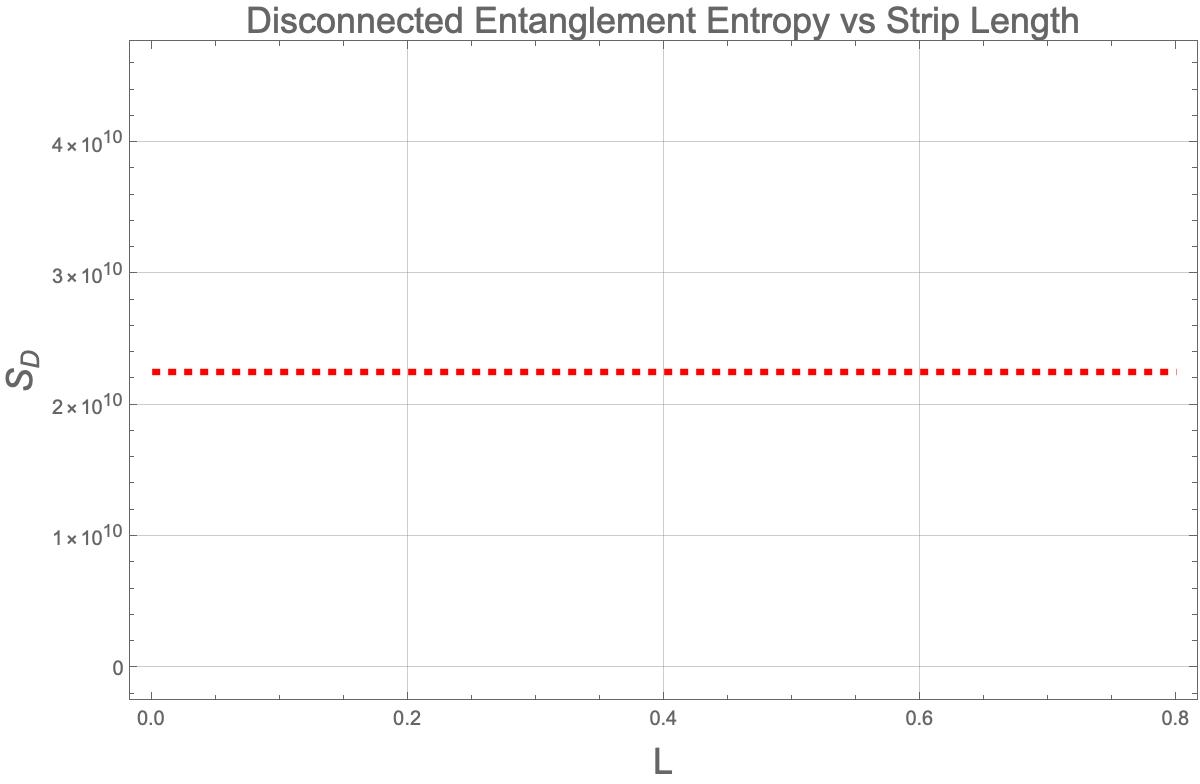} \ \ 
\includegraphics[width=0.43\textwidth]{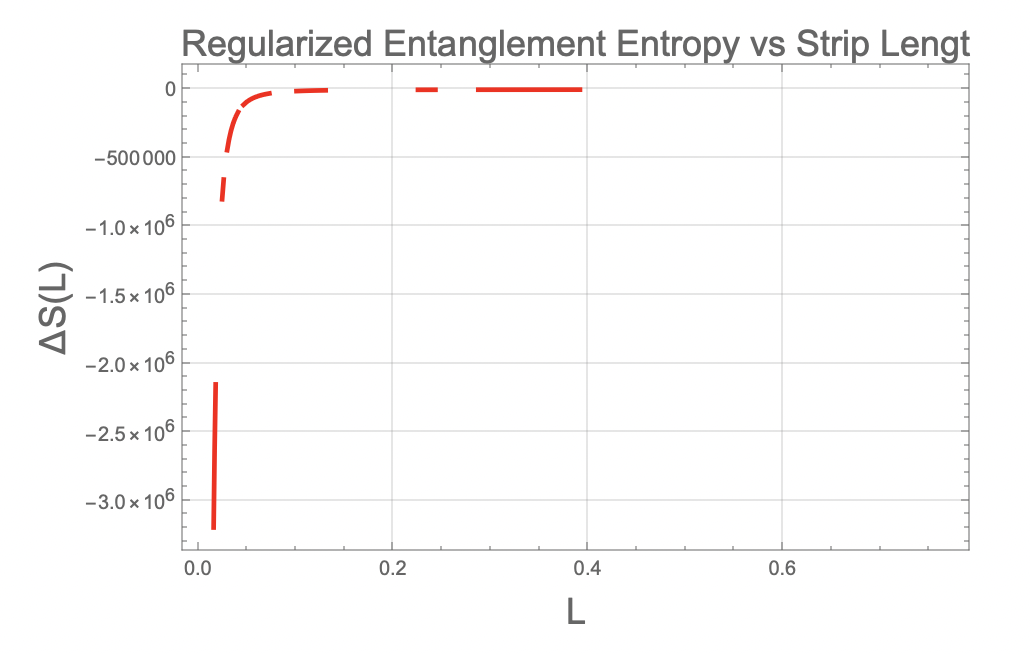} 
\caption{The behavior of $L(\rho_0)$ versus turning point $\rho_0$ for the wormhole metric \ref{eq:type2worm}, on the upper left part, the behavior of the function $S_C({\rho_0})$ versus $L(\rho_0)$ for the wormhole metric in the upper right, the behavior of the function $S_D({\rho_0})$ versus $L$ in the lower left part, and the behavior of $\Delta S(L)$ versus $L$ in the lower right are shown. Here, $b=3$, $\beta_1=1$, $\beta_2=0.5$, $g_s=1$, $G_N=1$, and $\rho_{\text{max}}=5$.}
\label{fig:SLwormsmallb}
\end{center}
\end{figure}

For the real values of $\beta_1$ and $\beta_2$, we can now postulate that the function $\beta(\rho)$ should be \textit{decreasing}, similar to the cases of confining backgrounds, which from Fig. \ref{fig:betawormhole3}, one could see that is the case. Note that as found in \cite{Kinar:1998vq}, since $\alpha(\rho)$ has a minimum and $\alpha\sqrt{\beta}$ diverges, this wormhole metric, admits linear confinement. This can also be seen from the plot of $\Delta S$ versus $L$, as in Fig. \ref{fig:SLwormsmallb}. However, this line here in this case, unlike other usual confining geometries, would be a flat line with zero slope.

Next, the function $H(\rho)= e^{-4\Phi} V_{\text{int}}^2 \alpha^d$ could be constructed, where $V_{\text{int}} = \int d\vec{\theta} \sqrt{\text{det} \lbrack g_{ij} \rbrack }$.  Since the term $dx^\mu dx_\mu$ has only one spatial component, effectively $d=1$, so we have
\begin{gather}
H(\rho)= \frac{\pi^{10} L^8 b^4}{16 g_s^4} (b^2 \cosh(2\rho) -1)^2 (\beta_1 e^{2\rho} + \beta_2 e^{-2\rho} )^2,
\end{gather}
which, as seen from the right part of Fig. \ref{fig:betawormhole3}, is a monotonically \textit{increasing} function as one would expect. In fact in $\rho \to \infty$, we have $H(\rho) \sim e^{8\rho}$. Also, when $\rho \to 0$, we have 
\begin{gather}
H(0) \approx \frac{\pi^10 b^4}{16 g_s^4} (b^2-1)^2 (\beta_1 + \beta_2)^2,
\end{gather}
which is finite and positive, and therefore consistent with the wormhole throat at $\rho=0$.

 Therefore, based on these observations, we would expect that RG flows could be constructed in the wormhole geometries in the same way as well.

Note that, for real values of $\beta_1$ and $\beta_2$, the function $\beta(\rho)$ here should typically be a monotonically decreasing function.

Then, we can proceed to check the behavior of entanglement entropy and other correlation measures in these backgrounds to find universalities. The entanglement entropy for a connected minimal surface with the ending at $\rho_0$ would be
\begin{gather}
S_C(\rho_0) = \frac{V_{d-1} }{2 G_N^{(10)} } \int_{\rho_0}^\infty d\rho \sqrt{\frac{\beta(\rho) H(\rho) }{1- \frac{H(\rho_0) }{H(\rho) } } }. 
\end{gather}

The length of a line segment for the connected solution would also be
\begin{gather}
L(\rho_0)= 2 \int_{\rho_0}^\infty d\rho \sqrt{\frac{\beta(\rho) }{\frac{H(\rho) }{H(\rho_0)}-1} }.
\end{gather}

For the case of $V_{d-1}=V_0=1$, we get
\begin{gather}
S_C(\rho_0) = \frac{1}{2G_N^{(10)}} \sqrt{\frac{\pi^{10}}{2g_s^4} } \int_{\rho_0}^\infty d\rho \frac{(b^2 \cosh(2\rho) -1)^{3/2} }{\sqrt{1- \frac{(b^2 \cosh(2\rho_0) -1)^2 (\beta_1 e^{2\rho_0} + \beta_2 e^{-2 \rho_0} )^2 }{(b^2 \cosh(2\rho)-1)^2 (\beta_1 e^{2\rho} + \beta_2 e^{-2 \rho})^2} } },
\end{gather}
or 
\begin{gather}
S_C(\rho_0) = \frac{\pi^5}{g_s^2 \sqrt{2} G_N^{(10)}} \int_{\rho_0}^\infty d\rho \frac{(b^2 \cosh(2\rho) -1)^{3/2} }{\sqrt{1- \left(\frac{H(\rho_0)}{H(\rho)}  \right) } }.
\end{gather}

For regularizing, we should subtract the disconnected part, 
\begin{gather}
S_D = \frac{V_{d-1}}{2G_N^{(10)}} \int_{\rho_\Lambda}^\infty d\rho \sqrt{\beta(\rho) H(\rho)}.
\end{gather}

So we get
\begin{gather}
S_D = \frac{1}{2 G_N^{(10)} } \sqrt{\frac{\pi^{10}}{2 g_s^4} } \int_{\rho_\Lambda}^ \infty d\rho (b^2 \cosh(2\rho)-1)^{3/2}.
\end{gather}

So the regularized entropy would be
\begin{gather}
S(\rho_0) = \frac{1}{2G_N^{(10)}} \sqrt{\frac{\pi^{10}}{2g_s^4}}\left \lbrack \int_{\rho_0}^\infty d\rho \frac{(b^2 \cosh(2\rho) -1)^{3/2}}{\sqrt{1-H(\rho_0)/H(\rho)}} - \int_{\rho_\Lambda}^\infty d\rho (b^2 \cosh(2\rho)-1)^{3/2} \right \rbrack.
\end{gather}

However, for plotting it numerically, we should change this function a little bit and split $S_D$ at a fixed point $\rho_*$ as 
\begin{gather}
\Delta S(\rho_0) = \frac{1}{2G_N} \left \lbrack \int_{\rho_0}^{\rho_*} \frac{\sqrt{\beta H} }{\sqrt{1-H_0/H} } d\rho + \int_{\rho_*}^{\rho_{\text{max}}} \left ( \frac{\sqrt{\beta H} }{\sqrt{1-H_0/H} } - \sqrt{\beta H} \right) d\rho \right \rbrack - \frac{1}{2G_N} \int_0^{\rho_*} \sqrt{\beta H} d\rho.
\end{gather}
One should choose a small enough $\rho_*$ that the second integrand is negligible for $\rho > \rho_*$, but large enough that the first integral is numerically well-behaved.

The length $L(\rho_0)$ in KKM also could be found from
\begin{gather}
L(\rho_0)= 2 \int_{\rho_0}^ \infty d\rho \sqrt{\frac{\beta(\rho) H(\rho_0) }{H(\rho)-H(\rho_0)} }.
\end{gather}

The behavior of $L(\rho_0)$ versus turning point $\rho_0$ for the wormhole metric \ref{eq:type2worm}, and the behavior of connected, disconnected and regularized entanglement entropy versus width of the strip are shown in Figure \ref{fig:SLwormsmallb}. From this figure, it could be seen that by decreasing $\beta$, the constant value of entanglement entropy would increase, which is expected.

By changing the parameters of the metric, one could see that increasing $b$ which is the throat of the wormhole, would make $L(\rho_0)$ smaller at each specific $\rho_0$, but makes the absolute value of regularized entanglement entropy $\Delta S(L)$ bigger for each $L$. 

\begin{figure}[ht!]   
\begin{center}
\includegraphics[width=0.9\textwidth]{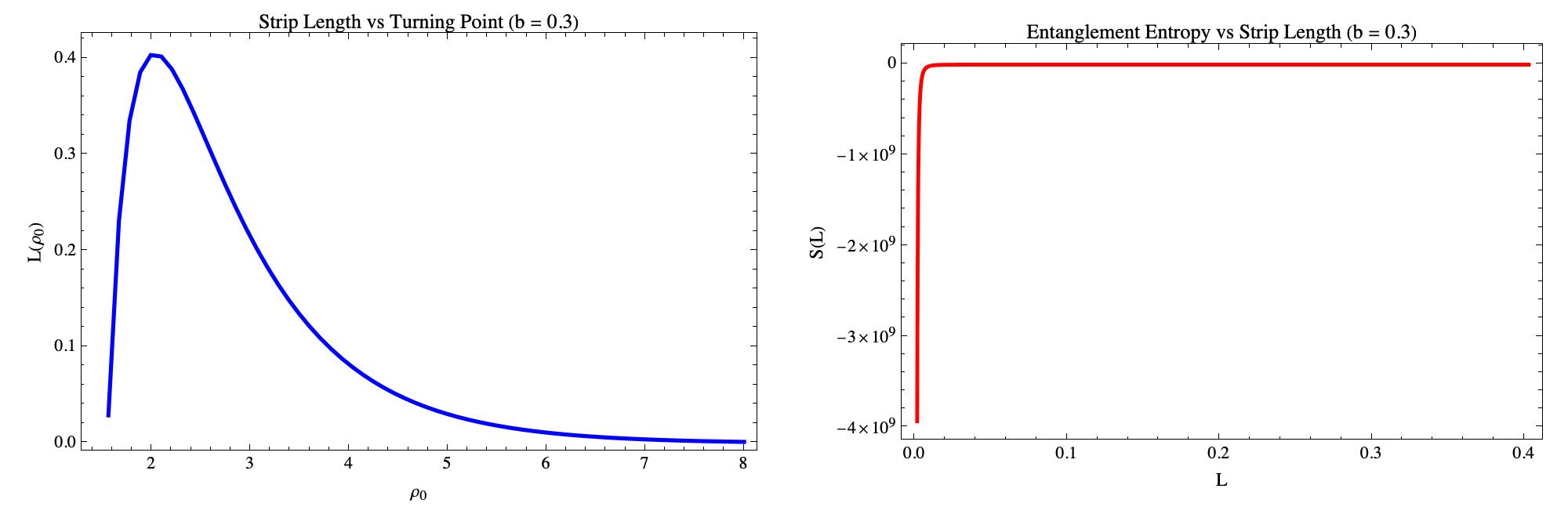}  
\caption{The behavior of $L(\rho_0)$ versus turning point $\rho_0$ for the wormhole metric \ref{eq:type2worm}, and the behavior of regularized entanglement entropy $S(L)$ versus $L$. Here, $b=0.3$, $\beta_1=1$, $\beta_2=0.5$, $g_s=1$, $G_N=1$, and $\rho_{\text{max}}=5$. Since $b<1$, we get a clear phase transition which is different from the case in Fig. \ref{fig:SLwormsmallb}. }
\label{fig:SLwormbigb}
\end{center}
\end{figure}

Here, $\beta(\rho)$ also has a finite value at the origin $\rho=0$, and therefore as explained in \cite{Kol:2014nqa}, the condition for having a phase transition is being met. However for the case of $b>1$, we do not see a clear butterfly phase transition in the behavior of entanglement entropy which as shown in \cite{Kol:2014nqa} which is a sign of non-locality. 

In order to get a clear phase transition and a butterfly phase diagram, the throat of the wormhole should be small enough and indeed we should have the condition $| b | <1$.

 \begin{figure}[ht!]   
\begin{center}
\includegraphics[width=0.9\textwidth]{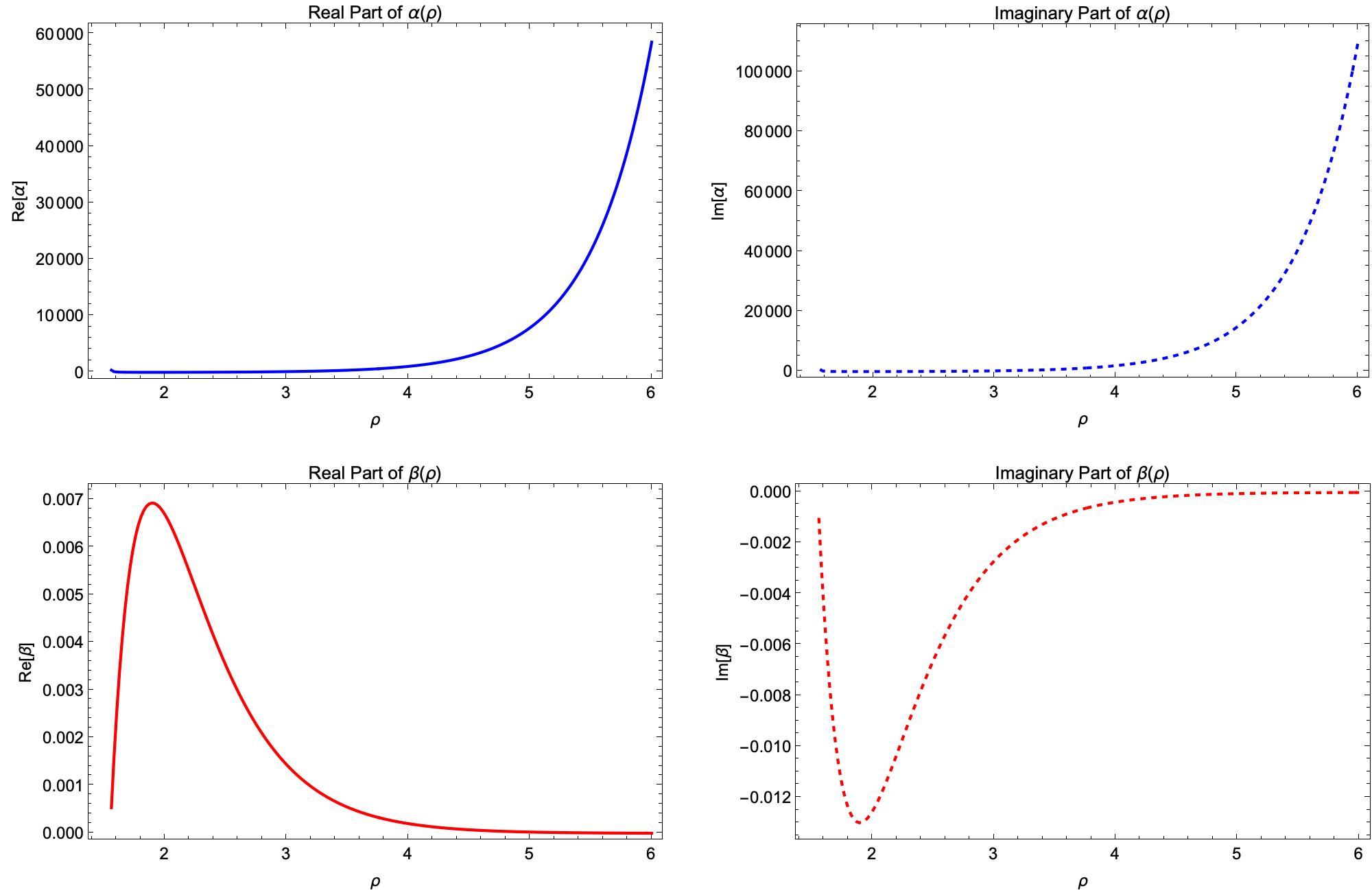} 
\caption{The behavior of the real and imaginary parts of the functions $\alpha(\rho)$ and $\beta(\rho)$ versus $\rho$ for the case of $\beta_1=5+3i$, $\beta_2=5-3i$, $b=0.3$, $g_s=G_N=1$ is shown. }
\label{fig:alphabetawormF}
\end{center}
\end{figure}

 \begin{figure}[ht!]   
\begin{center}
\includegraphics[width=0.9\textwidth]{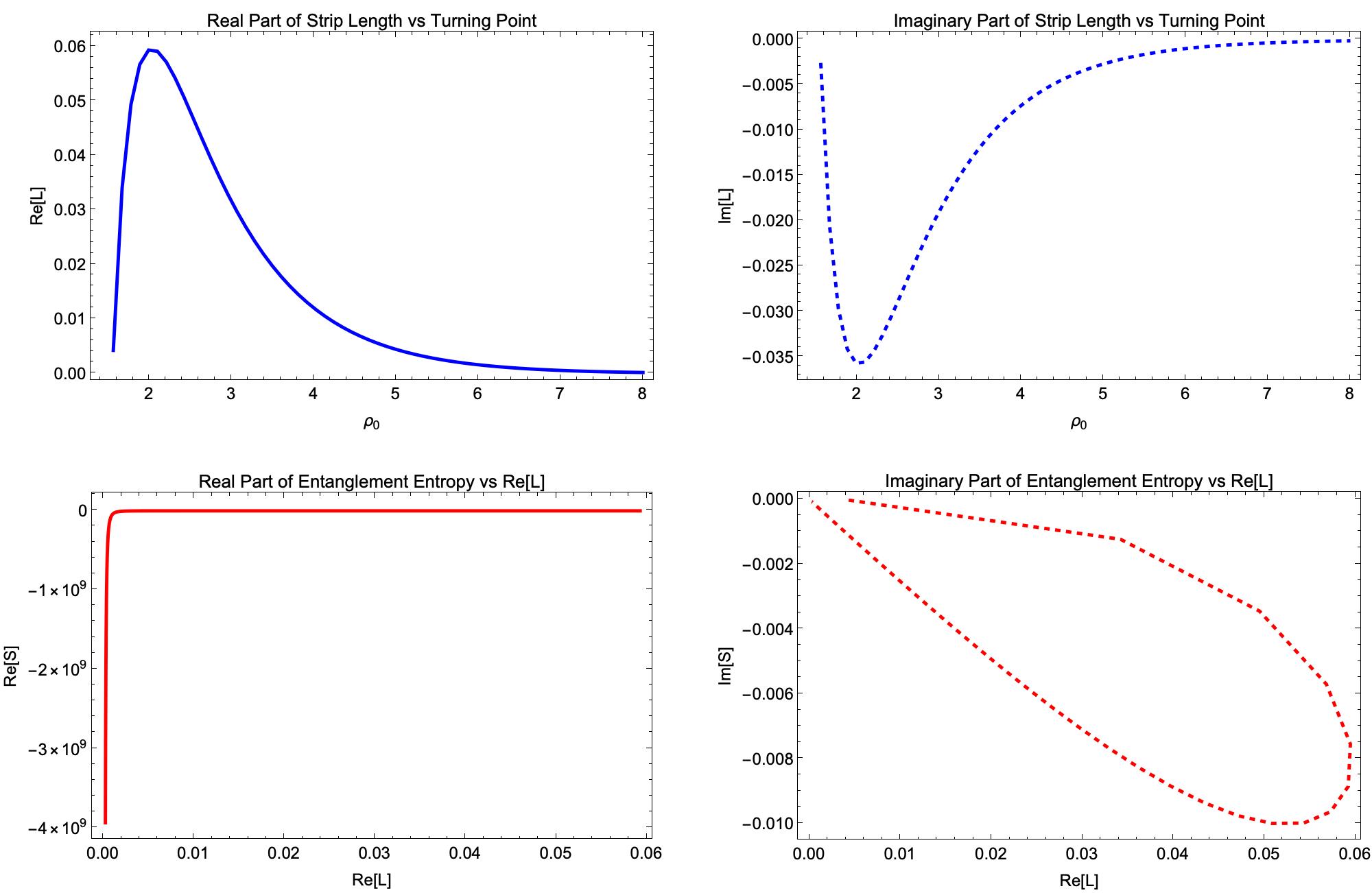} 
\caption{The behavior of the real and imaginary parts of strip length $L$ versus turning point and entanglement entropy versus $L$ for the case of $\beta_1=5+3i$, $\beta_2=5-3i$, $b=0.3$, $g_s=G_N=1$ is shown. }
\label{fig:alphabetawormS}
\end{center}
\end{figure}

 \begin{figure}[ht!]   
\begin{center}
\includegraphics[width=0.9\textwidth]{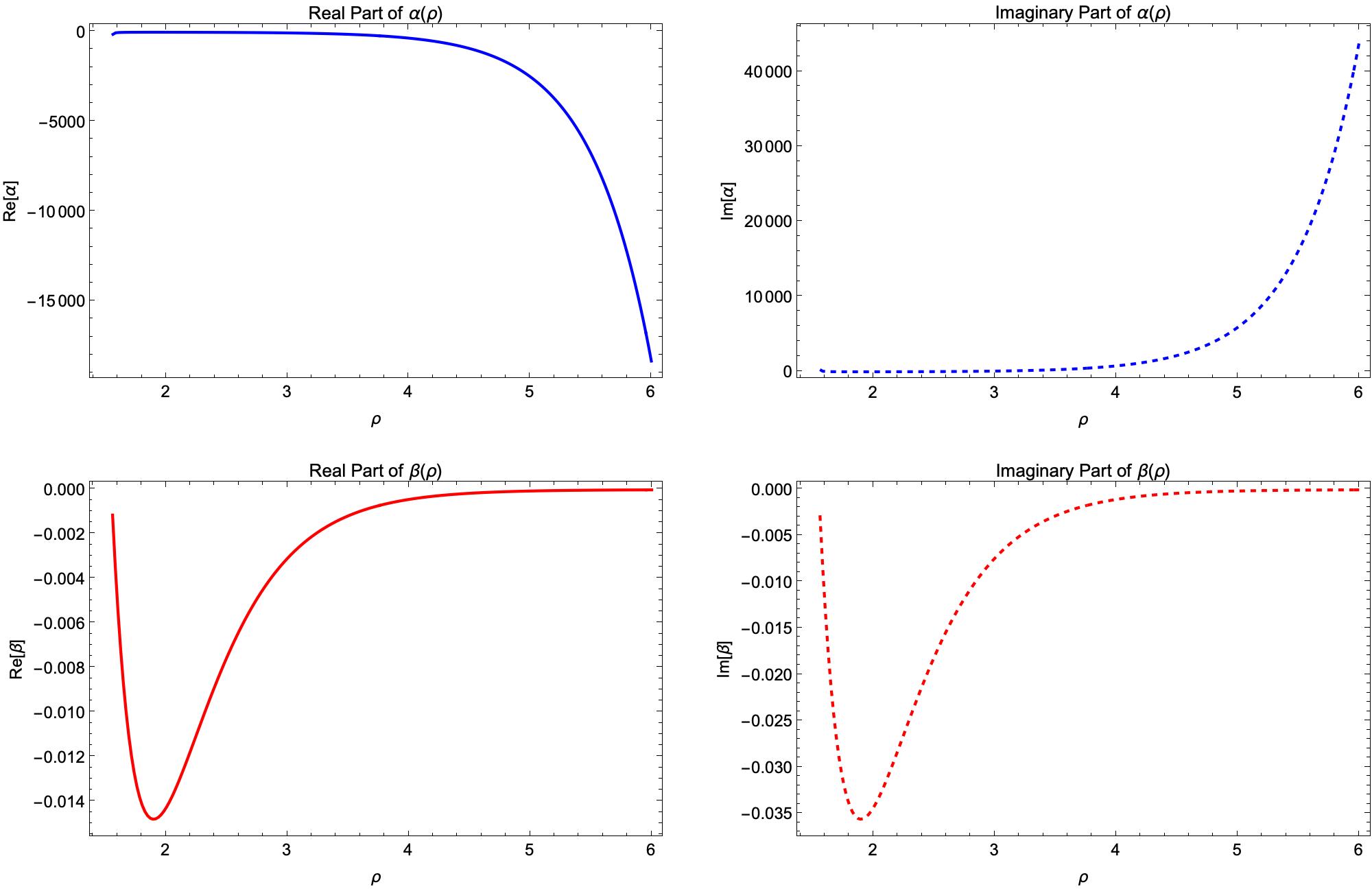} 
\caption{The behavior of the real and imaginary parts of the functions $\alpha(\rho)$ and $\beta(\rho)$ versus $\rho$ for the case of $\beta_1=2+3i$, $\beta_2=2-3i$, $b=0.3$, $g_s=G_N=1$ is shown. }
\label{fig:alphabetawormF2}
\end{center}
\end{figure}

 \begin{figure}[ht!]   
\begin{center}
\includegraphics[width=0.9\textwidth]{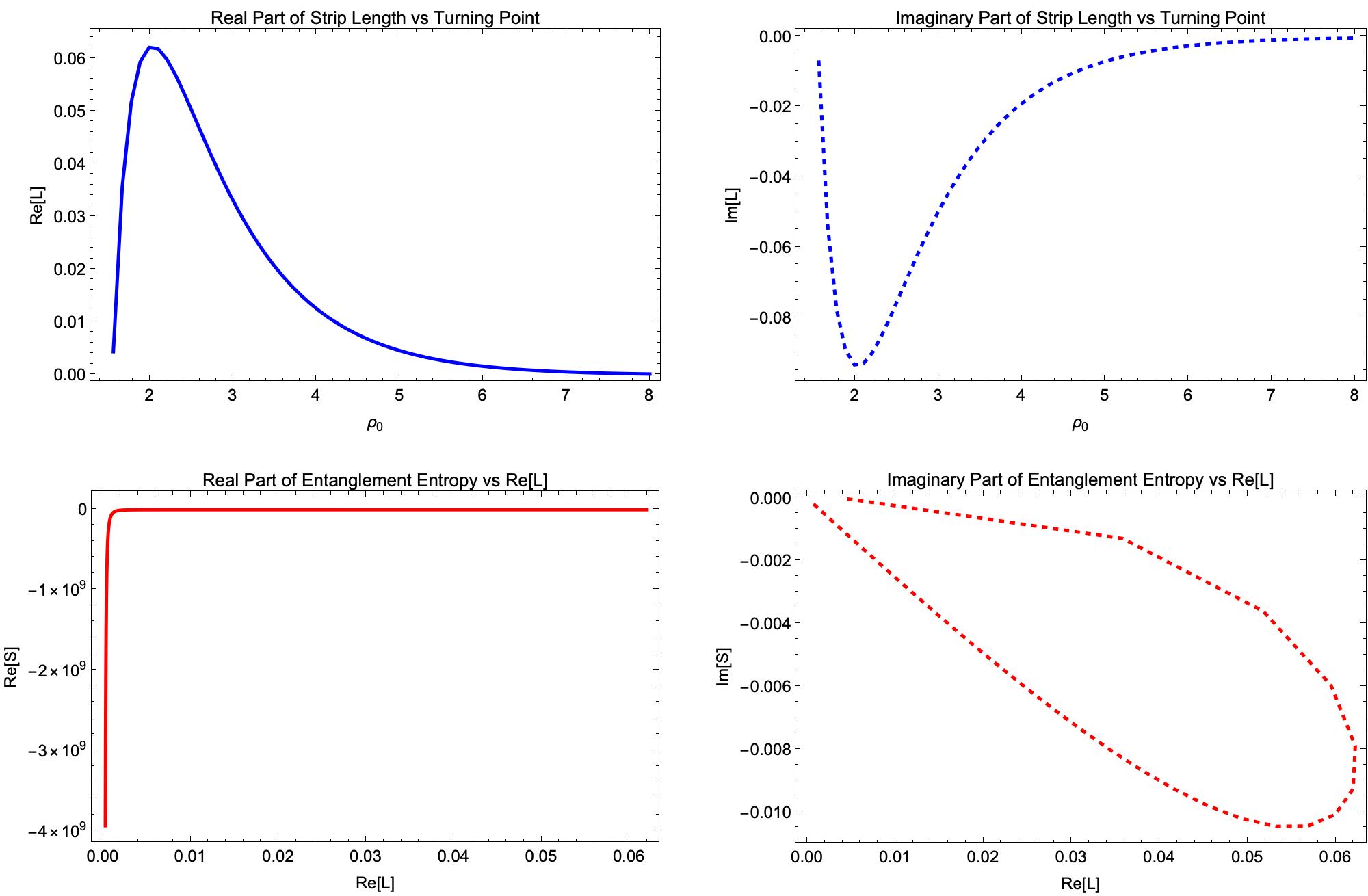} 
\caption{The behavior of the real and imaginary parts of strip length $L$ versus turning point and entanglement entropy versus $L$ for the case of $\beta_1=2+3i$, $\beta_2=2-3i$, $b=0.3$, $g_s=G_N=1$ is shown. }
\label{fig:alphabetawormS2}
\end{center}
\end{figure}

In Fig. \ref{fig:SLwormbigb}, the behavior of strip versus turning point, and also entanglement entropy versus the size of strip, for the case of small size of wormhole, i.e, $|b| <1$ is shown. This is again expected for the wormhole solution as it is directly connected to non-locality in the setup. Note that in the case of this wormhole solution also the two conditions of \cite{Kol:2014nqa} are being satisfied. For the IR condition, $\beta(\rho)$ is being diverged slower than $1/(\rho-\rho_\Lambda)^2$ near the horizon, and $\beta(\rho)$ decays faster than $1/\rho^2$ as $\rho \to \infty$.

Note that for the case of confining geometries, when this study is being done for a Wilson loop, we get a linear function in $S$ after a certain point which in this case has a zero slope.

Now for various values of $\beta_i$ ($\beta_1$ and $\beta_2$) ,we can study the behavior of $\alpha(\rho)$ and $\beta(\rho)$. If we extend the case to imaginary values for $\beta_i$, our metric changes from a Euclidean wormhole to a Lorentzian ``double cone" geometry. Here, $\beta_1$ and $\beta_2$ are the circumference of the Euclidean time circles on the two asymptotic boundaries. For the case of $\beta_1=iT$, $\beta_2 = -iT$, and $T$ real, we get a double cone geometry of \cite{Saad:2018bqo}.

One could see from these figures that, if the imaginary part is smaller than the real part, as in Figures \ref{fig:alphabetawormF} and \ref{fig:alphabetawormS}, both the real parts of $\alpha$ and the imaginary part are monotonically increasing. In this case, the real part of $\beta(\rho)$ has a maximum and the imaginary part of $\beta(\rho)$ has a minimum.

On the other hand, if the imaginary part is bigger than the real part, as in Figures \ref{fig:alphabetawormF2} and \ref{fig:alphabetawormS2}, the real part of $\alpha$ is monotonically decreasing and the imaginary part is monotonically increasing. Also in this case, both the real and imaginary parts of $\beta(\rho)$ have a minimum. Note that physically $\alpha(\rho)$ is the size of Euclidean time circle and $\beta(\rho)$ which is its inverse is the local inverse temperature. So, when the real part of $\beta_i$ is bigger, the geometry is more like a Euclidean wormhole where $c$-theorem forces the time circle to grow and temperature to decrease from the bottleneck to the boundary, but with a maximum in the bottleneck where the temperature is the highest. On the other hand, when the imaginary part is bigger, the geometry is effectively Lorentzian, where the size of the time circle becomes a phase and Re$\lbrack \alpha \rbrack$ decreases as the phase rotates.

Note that the plot of $\text{Im} (S)$ versus $\text{Re} (L)$ is exactly similar to Figure. 2 of \cite{Ali-Akbari:2026dtl}, where the imaginary part of holographic timelike entanglement entropy (HTEE) as a function of $\mu T^{-1}$ is shown. There is only  a minus sign difference which makes it flipped over the x-axis. There, the unstable and stable regimes were also specified.
Also in \cite{Ali-Akbari:2026dtl}, the authors discussed how the imaginary part of pseudoentropy is tied to the coherence between the initial and final states. It is only non-zero when there exists quantum coherence between the two states. In the case of the wormhole, when the real parts or the imaginary parts of both $\beta_I$ are zero, we get a vanishing imaginary entanglement entropy. In fact, when both the real and imaginary parts of $\beta_I$ become zero, the two sides of wormhole get disconnected and there would be no coherence between them. Also, when $b>1$, again we see the same behavior, as the coherence would be lost.

They also found a universality as
\begin{gather}
\left( \frac{\partial \, \text{Im}[s_{\Delta t}]}{\partial(\mu/T)} \right)_T \sim \left[ \left( \frac{\mu}{T} \right)_c - \frac{\mu}{T} \right]^{-\theta_I}, \qquad \boxed{\theta_I = \frac{1}{2}}.
\end{gather}
where the exponent $\theta_I =1/2$ would be probe-independent and is just a property of the background geometry rather than a particular probe. Therefore, this universality is also related to the partition functions, competitions between saddles, and the universalities that we discussed in the previous section.

So by checking these behaviors for wormhole solutions and comparing with various solutions of confining geometries of supergravity models, one can detect new universalities.

\subsection{Level statistics in wormholes}
Now we can study level statistics and RG flows in wormhole soutions. As for the precision test, the example of a dilaton interpolating between the two values of Yang-Mills coupling $g_{\text{YM}}$ \cite{Cotler:2022rud}, which there describes the level statistics , can be written as
\begin{gather}
\langle \mathrm{tr} ( e^{- \beta_1 H(g_{\text{YM}} ) } ) \mathrm{tr}( e^{- \beta_2 H(g_{\text{YM}} + \delta g_{\text{YM}} ) } ) \rangle.
\end{gather}

After analytically continuing in terms of $\beta$, the result could be written as
\begin{gather}
\langle \mathrm{tr} ( e^{- ( \beta + i T)  H(g_{\text{YM}} ) } \mathrm{tr} ( e^{- (\beta- i T)  H(g_{\text{YM}} + \delta g_{YM}  ) } \rangle_{\mathcal{C}}  \sim T e^{- \beta E_{\text{tot}}  - \frac{\pi N^2}{32} \sqrt{\frac{E_{\text{tot} } }{2E_0}  }   \left( \sqrt{\frac{E_{\text{tot} } }{2E_0} } -1  \right ) \delta g_{\text{YM}}^2 T }. 
\end{gather}

Note that in the bulk holographic picture, the dilaton interpolates between the two values of the two boundary CFTs.

\begin{figure}[ht!]   
\begin{center}
\includegraphics[width=0.45\textwidth]{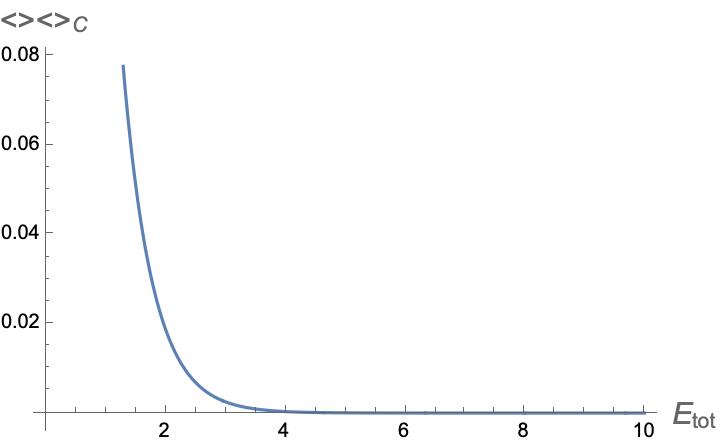} 
\includegraphics[width=0.45\textwidth]{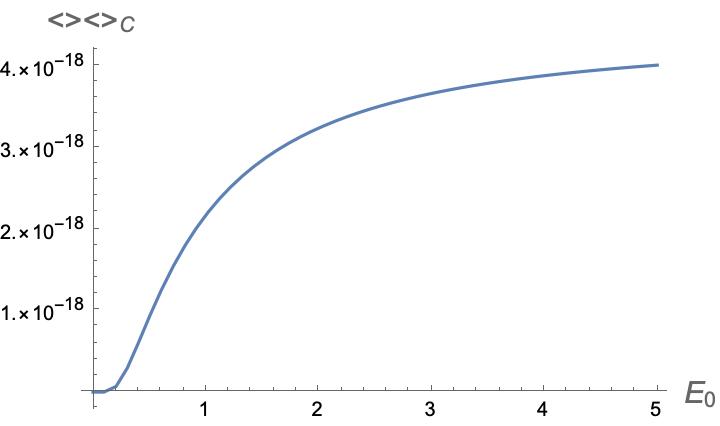}  
\includegraphics[width=0.45\textwidth]{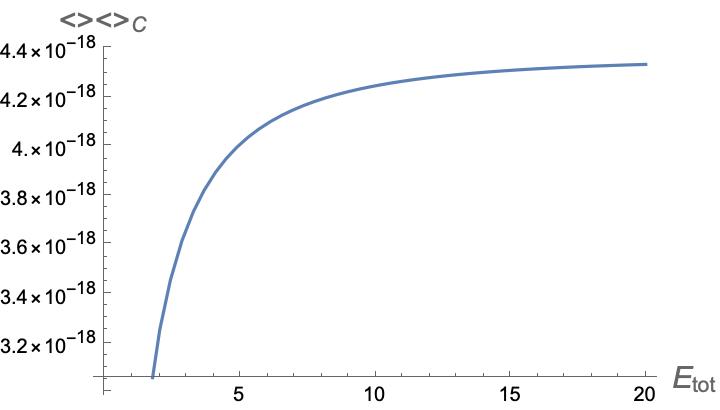}   
\caption{The behaviors of level statistics (LS) between two CFTs with different couplings, $g_{\text{YM} }$ and $g_{\text{YM}} + \delta g_\text{YM}$, which match with the c-flows across theories with different dimensions.}
\label{fig:flowsfrompartition1}
\end{center}
\end{figure}

\begin{figure}[ht!]   
\begin{center}
\includegraphics[width=0.5\textwidth]{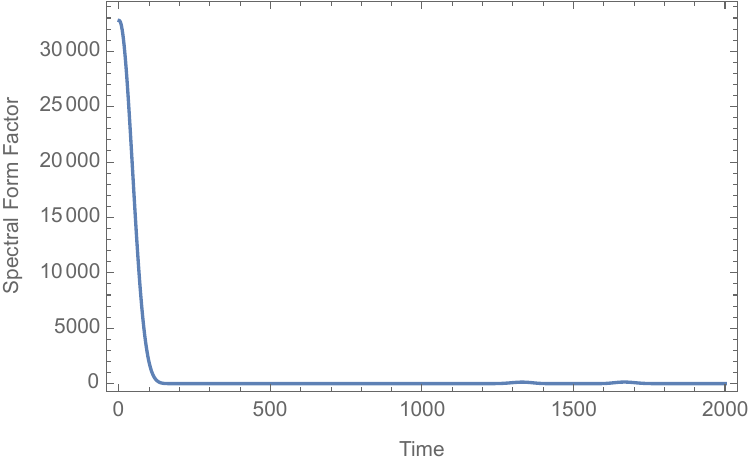} 
\caption{The behavior of spectral form factor for $N=10$ and $q=2$ is shown.}
\label{fig:lowNSFF}
\end{center}
\end{figure}

The behavior of this correlation with respect to $E_0$ and $E_{\text{tot}}$ is shown in Fig. \ref{fig:flowsfrompartition1}, which as one could see, they behave monotonically and their behavior matches the behavior of holographic c-functions between dimensions as found in \cite{GonzalezLezcano:2022mcd}. Therefore, this two-point function can be considered as a flow as well. So by using this level statistics (LS), we can define the ``LS-theorem". One then could check if this theorem as an example of a ``g-theorem" can survive after coupling to higher dimensional CFTs, bringing other sectors of theories or adding different couplings. In fact, we can note that the spectral form factor for some regimes could also behave as a c-function as seen in Figure \ref{fig:lowNSFF}.

We can also analyze supersymmetric index for the case of wormholes. One should note that the supersymmetric index is known to be independent of RG flow, but the partition functions and level statistics can indeed define a RG flow.

Another piece of evidence that the wormhole structures can lead to new RG flows comes from the smoothness of the ramp in the spectral form factor which is due to the long-range repulsions in the spectrum of the boundary Hamiltonian. These repulsions in the spectrum of microstates, which can be found from the wormhole amplitudes can also be confirmed using the connections between RG flow and the velocity of the island, as discussed in \cite{Ghodrati:2022hbb}. 

So, due to the level repulsions between two CFTs across different dimensions and also across different couplings, their eigenvalues are mutually exponentially de-correlated.

In the case of de-correlation across dimensions, the difference between dimensions can be included. Here, $D-d$ can determine the slope and the strength of de-correlation, as observed in \cite{GonzalezLezcano:2022mcd}, since the behavior of the local holographic (LH) c-function would be
\begin{gather}
c_{LH} (z) = \frac{\ell^{D-d} }{ ( (e^{-\tilde{f}(z)} )' )^{d-1} },
\end{gather}
where $\tilde{f} (z)$ is the effective IR warp factor
\begin{gather}
\tilde{f} (z) \equiv f(z) + \frac{D-d}{d-1} g(z).
\end{gather}

So the results of \cite{GonzalezLezcano:2022mcd} should match the predictions of the parametric random matrix theory of  Altshuler and Simmons \cite{Simons:1993xs, Szafer:1993zz, Kravtsov:1994zz, Simons:1993zza}.

Another important example studied in \cite{Cotler:2022rud} is in a D3 brane with a unit of RR flux which corresponds to the correlations between $SU(N)$ SYM and $SU(N+1)$ SYM. The result found in \cite{Cotler:2022rud} is
\begin{gather}
\langle \mathrm{tr} ( e^{-(\beta+ i T )H_N } ) \mathrm{tr} (e^{-(\beta-i T) H_{N+1} }) \rangle_{\mathcal{C}} \sim T e^{-\beta E_{\text{tot}} - N T \frac{3}{2} i^{1/2}  (\frac{E_{\text{tot}} }{2N} )^{2/3} (1+O  \left(( \frac{ E_{\text{tot} }  }{2N})^{-1/3}  )\right)},
\end{gather}
which can then be tested by SUSY sigma-model method. This can also be extended to $SU(N+M)$ and again we expect the difference, i.e., $M$ determines the strength of the de-correlation. 

Then, in these examples, the notion of smearing of energy, and smearing of energy eigenstates, from the point of view of c-functions across dimensions, similar to \cite{GonzalezLezcano:2022mcd} could be studied. This smearing over energies of a chaotic system was first studied by Berry \cite{1989RSPSA.423..219B} and Heller \cite{Heller:1984zz}, et al. in  the 70s and 80s. So for the full quantum gravity, we can also do smearing across dimensions, using partial dimensional reduction and coarse graining.

As we discussed, using entanglement structures and RG flows, in this regard, we can expect to see connections with island formalism as well. In fact, in \cite{Matsuo:2024ypr}, the universalities of islands have also been discussed. The authors showed that the island is outside the horizon for stationary black holes and is inside of horizon for evaporating black holes. The connections between these behaviors of islands, and the universalities in the behavior of partition functions themselves, can then be investigated. The contributions and effects of Hartle-Hawking states and brane states on these behaviors could then be separated.

An interesting follow-up idea would be to connect optimal transport equations and wormhole structures. Note that the connections between RG flow equations and optimal transport has been discussed in \cite{Cotler:2022fze}, and the connections between RG flow and wormholes have been discussed in \cite{Ghodsi:2022umc}. Now the problem is to connect wormholes and optimal transport themselves. A key ingredient would be the Wasserstein-2 distance $\mathcal{W}$ which is defined on the space of probability distributions, which is in the form of
\begin{gather}
\mathcal{W}_2 (p_1, p_2) := \left( \underset{{\pi \in \Gamma(p_1,p_2)}}{\mathrm{inf}}  \int_{M \times M} dx \  dy  \ \pi(x,y) |x-y|^2  \right)^{1/2},
\end{gather}
where $\Gamma(p_1, p_2)$ is the space of probability distributions $\pi (x,y)$. Note that in these calculations, we have the constraints $\int_M dy \ \pi(x,y)= p(x)$ and $\int_M dx \  \pi(x,y) = q(y)$. 

In \cite{Cotler:2022fze}, it has been shown that the Polchinski equation or more generally the Wegner-Morris equation would be the optimal transport gradient flow of a relative entropy. Therefore, it would be logical to think that wormhole solutions found in \cite{Ghodsi:2022umc} for the case of RG flows could also be related to the optimal transport flow of some relative entropy as well. Therefore, RG flows, wormholes, optimal transport equations and relative entropy could all fit in a connected structure.

Another important quantity in studying wormholes, their dynamics, and their connections to chaos and quantum correlation measures is the analytically continued partition function $| Z(\beta + it |^2$, or more specifically
\begin{gather}
g(\beta, t) = \frac { \Big \langle | \mathbb{Z} ( \beta+ i t ) |^2\Big \rangle_J }{ \langle \mathbb{Z} (\beta) \rangle^2_ J}.
\end{gather}

Note also that wormhole structures would cause a smooth ramp over a long timescale in the spectral form factor of $\langle \mathrm{tr} ( e^{-(\beta+ i T)H } ) \mathrm{tr} (e^{-( \beta- i T)H } ) \rangle$, where this ramp is produced as a result of the long-range repulsion in the spectrum of the boundary Hamiltonian. The spectral form factor of the SYK model has a universal behavior with four mesoscopic regimes, namely ``ballistic", ``diffusive" (slope), ``ergodic" (ramp),  and ``quantum" (plateau) regimes \cite{Cotler:2016fpe, Liu:2016rdi}, as shown in Figure \ref{fig:SFFphases}. The ramp here corresponds to the ergodic regime. The footprints of these four saddles on the structure of partition function of chaotic systems are universal and could be analyzed further. We did some initial studies in \cite{Ghodrati:2022hbb}.

The random matrix behavior observed in \cite{Cotler:2016fpe}, and the four stages of the dynamics found through the spectral form factor can also be observed using other mixed correlation measures. This could be done by fixing time and changing the size of the subsystems, as observed in \cite{Ghodrati:2021ozc,Ghodrati:2022hbb,Ghodrati:2022kuk,Ghodrati:2020vzm}, 
which were studied for QCD supergravity confining backgrounds. These measures were complexity, complexity of purification, entanglement of purification and mutual information. 

In \cite{An:2023dmo}, the replica wormhole has been considered as a vacuum-to-vacuum transition, which is again in line with the picture of RG flows in field theories. They proposed that the corresponding gravitational partition function would be controlled by the manifold of these degenerate vacua.

The partition function of an $n$-folded wormhole is \cite{Hirshfeld:1999xm}
\begin{gather}
Z(\Sigma_n) = \sum_j \frac{e^{-n \beta E_j} \chi_j(g_\beta)^n } {d_j^{n-2} },
\end{gather}
where for the case $\chi_j(g_\beta) = z_j (\beta)$, it becomes $Z_j(\Sigma_n) = d_j^{2-n} = \text{Pf}(\bar{\omega})^{2-n}$ \cite{Verlinde:2021kgt}.
This also could be regarded as a particle moving on a symmetric space such as in the presence of a magnetic field, with the partition function of
\begin{gather}
Z(\beta) = \int \lbrack dX \rbrack e^{\int_D \Omega - \oint_{\partial D} H dt }.
\end{gather}

There are other interesting supergravity models where the RG flows have been constructed, which would lead to many universalities, especially the universalities in the partition functions. For example, in \cite{Chen:2020mtv}, some solutions of $4d$, $\mathcal{N}=2$ gauged supergravity coupled to vector multiplets, which are holographically dual to superconformal line defects have been constructed.  Then, in \cite{Chen:2021mtn}, in $3d$ gauged supergravity with sixteen supersymmetries, a Janus-type solution has been constructed and by varying the supersymmetry parameters, RG-flow interfaces have been examined.  Then, in \cite{Gutperle:2022fma}, the BPS flow equations and holographic entanglement entropy have been calculated. The partition function for all these cases, their connections to the RG flows studied here, and the universalities could then be further explored.

As we discussed wormhole geometries, the entanglement entropy in their backgrounds, and their RG flows, it would be interesting to check their applications in other areas of quantum gravity. In fact, recently in \cite{Bah:2022uyz}, using a wormhole configuration, the square of the amplitude of a spherical shell has been calculated. In \cite{Balasubramanian:2022lnw}, it has been shown that quantum mechanical wormholes can create overlaps in the Hilbert space, which can explain the microscopic origin of the Bekenstein-Hawking black hole entropy. In \cite{Chen:2020tes}, the bra-ket wormholes in gravitationally prepared states have been studied. The Lorentzian interpretation of these bra-ket wormholes would be related to expanding and collapsing Friedmann cosmology, and the entropy of an interval in such geometries is related to de Sitter entropy.

In \cite{Ghodsi:2022umc}, wormholes connecting different QFTs with different couplings and the connections with holographic interfaces of flat-space QFTs have been discussed.  In addition, the phenomena of ``walking" flows and the generation of extra boundaries by ``fragmentation" of flows has also been discussed.

Also in \cite{Blommaert:2023vbz}, the Lorentzian wormholes have been studied further, where these wormholes contain  ``crotches" on their saddles located at an extremal surface. Their main result was that these Lorentzian wormhole geometries would be gauge-equivalent to path integrating over all ``mostly" Euclidean smooth spacetimes, which is connected to the story of RG and wormholes. So this path integral would be explained by an RG picture.

Another interesting question regarding wormholes is the structure of the quantum states of qubits passing through the wormholes, specifically their structure in the middle of the throat or at the minimal wedge cross section, versus their initial and final states, and what the mixed correlation measures such as mutual information or negativity could tell us about the evolution of quantum states. This question would specifically be related to the c-theorem, monotonicity and RG flows \cite{Jensen:2015swa}, where mutual information could be implemented, as in \cite{Byun:2026ewk}.

In the literature, there are different other ways of constructing wormholes, in string theory or in bottom-up setups, where various solutions have been discussed. For instance, in \cite{deBoer:2022rbn}, worldsheet traversable wormholes have been constructed by a new double-trace deformation of the Lagrangian which was different from the standard double-trace deformation of \cite{Gao:2016bin}. However, these deformations are not the only way of constructing wormholes, and other constructions are also possible.

In \cite{Azad:2023iju},  the radial perturbations of Ellis-Bronnikov wormholes ($l=0$)  in a slow-rotation expansion up to second order has been studied, and it reports indications that rotation may remove this particular instability. However, it does not establish complete mode stability, nor does it show that the result extends to other wormhole solutions. So their methods could be examined for other wormhole solutions and check whether by adding rotation to wormholes male them linearly stable. For these new wormholes with rotation which then could be stable this way, one could find entanglement entropy, RG flows, and universality classes.

It would also be interesting to investigate the effects of various parameters on this geometry using different quantum information measures. The metric is 
\begin{gather}
ds^2=-e^f \lbrack 1+ \epsilon_r^2 2 ( h_0 (r) + h_2 (r) P_2( \theta)) \rbrack dt^2 + e^{-f} \lbrack 1+ \epsilon_r^2 2 (b_0 (r)+b_2 (r) P_2 (\theta)) \rbrack dr^2 \nonumber\\
+ e^{-f} R^2 \lbrack 1+ \epsilon_r^2 2 ( k_0(r) + k_2 (r) P_2 ( \theta)) \rbrack \times \Big \{d\theta^2 + \sin^2 (\theta) \lbrack d\varphi-\epsilon_r w(r) dt^2   \Big \},
\end{gather}
where the slow-rotation parameter is $\epsilon \ll 1$ and $ R^2= (r^2 + r_0^2 )$, and the Legendre polynomial is $P_2( \theta)= ( 3\cos^2 (\theta) -1 )/2$. The second order phantom scalar field is
\begin{gather}
\Phi = \phi (r) + \epsilon_r^2 ( \phi_{20} (r) + \phi_{22} (r) P_2 (\theta) ).
\end{gather}

The throat of this wormhole is at $r=0$ with the area of $A=4\pi r_0^2 e^{\frac{\pi C}{2r_0}}$.

\begin{figure}[ht!]   
\begin{center}
\includegraphics[width=0.97\textwidth]{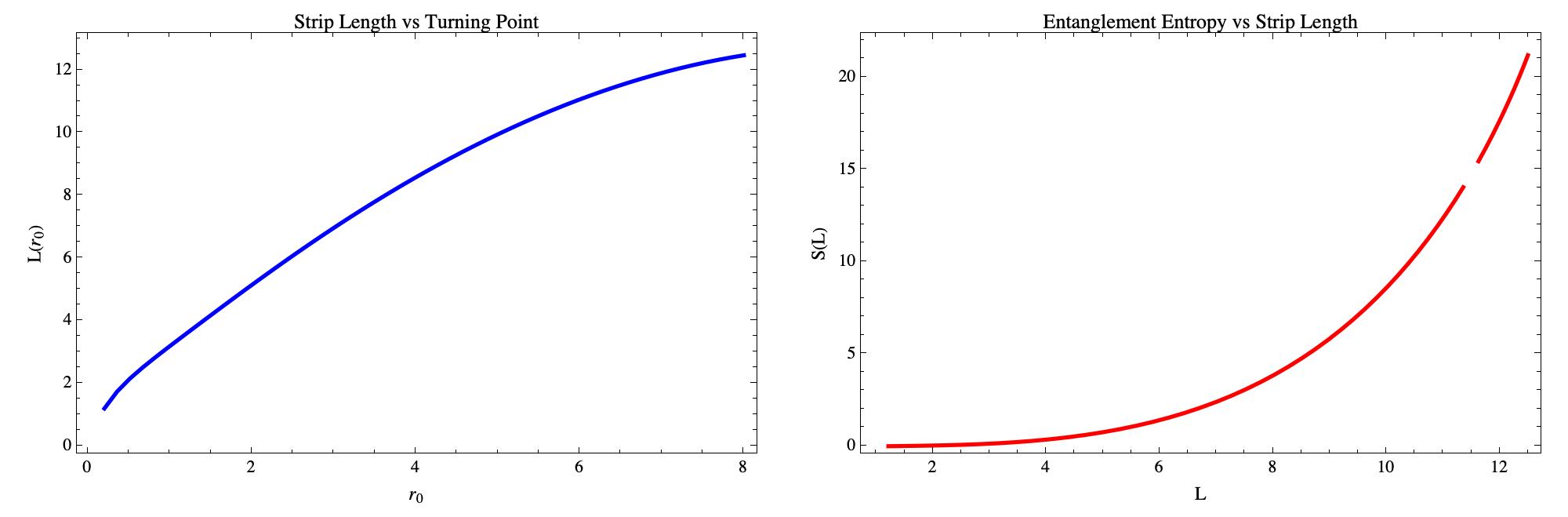} 
\caption{The behavior of width of a strip versus turning point and entanglement entropy versus $L$ for the Ellis-Bronnikov wormholes. }
\label{fig:rotEE}
\end{center}
\end{figure}

The dilaton field and the function $f$ in the background solution are as follows
\begin{gather}
f(r)= \frac{C}{r_0} \left ( \tan^{-1} \left ( \frac{r}{r_0} \right ) - \frac{\pi}{2} \right ), \ \ \ \phi(r) = \frac{Q f}{C}.
\end{gather}
 In the simpler symmetric background ($C=0$), the parameters are
 \begin{gather}
 \phi_{20}(r) = \frac{3J^2}{r r_0^3} \left ( \phi + \frac{\pi}{2} - \frac{3r}{2r_0}- \frac{4}{\pi r_0} - \frac{3r^2+ 4 r_0^2}{2 R^2 \phi}\right), \nonumber\\
 b_0(r) = \frac{\Delta M}{r} \left ( 1- \phi \frac{r_0}{r} \right) + \frac{6J^2}{r r_0^3} \left ( \frac{r_0}{r} + \phi \right) - \frac{3J^2}{r^2 r_0^2}\phi^2, \nonumber\\
 h_0(r) = \frac{3J^2}{r_0^4} \left (\phi^2 + \frac{\pi^2 - 8}{2\pi} \phi - \frac{r_0^2}{R^2} \right).
 \end{gather}
For this metric, the behavior of width of a strip versus turning point and entanglement entropy versus $L$ is shown in Fig. \ref{fig:rotEE}.

In \cite{Loges:2023ypl}, the $10d$ uplifted solution of Euclidean axion wormholes has also been discussed.  This uplift is from Giddings-Strominger wormholes in $\mathcal{N}=8$ Euclidean supergravity over a 6-torus to $10d$ type IIA supergravity. In addition, wormhole solutions in  AdS, massive type IIA on $S^3 \times S^2$, and in type IIB on $T^{1,1}$ have also been discussed.

In \cite{Loges:2023ypl} and in two of their references \cite{McNamara:2020uza,VanRiet:2020pcn}, it has been suggested that in $d>3$, wormholes could be spurious. However, we do not agree with this conjecture due to the universalities that we have observed in the phase structures of mixed correlation systems in various dimensions as in \cite{Ghodrati:2020vzm,Ghodrati:2021ozc,Ghodrati:2022kuk,Ghodrati:2022hbb}, specifically the case of phase diagrams coming from Logarithmic negativity in $2d$ JT as found in \cite{Dong:2021oad}, and the case of mutual information and critical distance between subsystems, as found in confining $10d$ supergravity models. In order to provide the required negative Euclidean energy-momentum tensor to stabilize and source the wormhole geometries, axion fields can then be used.

In \cite{Mirjalali:2022wrg}, the interior of black holes in massive gravity theories has been probed. When the correlation and entanglement is strong, which corresponds to a smaller potential,  the interior of the black hole would be a Kasner universe. By increasing the parameter $\alpha$ in equation 2.12 of \cite{Mirjalali:2022wrg}, the graviton mass would increase, the potential would become larger than the kinetic term, and the breaking of the diffeomorphism invariance suppresses the entanglement entropy and significantly changes the phase structure of the black hole interior.  This actually corresponds to the change of $u_{KK}$ in the $10d$ supergravity models we studied in \cite{Ghodrati:2021ozc}, which can induce transitions between saddles of entanglement entropy. Increasing $\alpha$ decreases the entanglement velocity $v_E$, as one would expect and as also observed in figure 9 of \cite{Mirjalali:2022wrg}. 

Now an interesting question is whether the entanglement velocity $v_E$, which satisfies the relation
\begin{gather}
\frac{dS}{dt(0)} = v_E \mathcal{V}_1 s, \ \ \ \ \ \ \ \ \ \ v_E^2= r^4_+ \frac{|f|  e^{-\mathcal{\chi}}}{r^4} |_{r=r_{\text{crit}} }
\end{gather}
and the butterfly velocity, which satisfies the relation 
\begin{gather}
v_B^2= r_+ \frac{ |f'| }{4} e^\chi |_{r=r_+}
\end{gather}
can probe the four main saddles we studied in \cite{Ghodrati:2021ozc}. Indeed, by examining figures 10 and 11 of  \cite{Mirjalali:2022wrg}, one can see that this is the case, and the four phases can be detected by changing the three main parameters of $v_E$, $v_B$ and $\phi_0$.

\subsection{Complexity and RG flows}

Complexity, as an interesting and useful quantum correlation measure, can also help us to find further universalities in the behavior of wormholes and the connections to the behavior of partition functions. For instance, in \cite{Zolfi:2023bdp}, using complexity = volume, the growth of complexity for a $2+1$ dimensional Lorentzian wormhole spacetime has been calculated. They found that complexity grows linearly at early times, but at later times, the growth would not be linear anymore, which actually matches with the general behavior of entanglement entropy and the functions we analyzed in section \ref{sec:RGflowsEE}. However, saturations at late times ($t \sim \exp(S) $) would still be satisfied due on the universal behavior of the partition functions in those limits.

In \cite{Erdmenger:2023wjg}, the connections between chaos and Krylov state complexity has been discussed. When the wavefunction spreads among the degrees of freedom of the two mixed systems, which can be measured by the Krylov state complexity, the four main universal phases, rise, slope, ramp, plateau of the mixed correlations behaviors, as in Figure \ref{fig:SFFphases}, and the transition probabilities could be detected where these transitions are driven by the normalized spectral form factor. The behavior of the spectral form factor (SFF) would be determined by the level statistics and level repulsions.

Note that in principle, there could be many RG flows, but one of them is actually the optimum. One way of specifying it is by using complexity, as the minimum complexity leads to the optimum flow. The RG flow for interacting QFTs using circuit complexity has actually been discussed in \cite{Bhattacharyya:2018bbv}.  Complexity can also be evaluated using the Liouville action, and the optimization procedure determines the most efficient way to represent the partition function. Circuit complexity can also be calculated for fractional dimensions, which touches on the idea of Wilson-Fisher fixed point in the epsilon expansion.

\section{Conclusion}

In this work, we discussed the universalities observed in black holes geometries as well as in various low-dimensional quantum gravity models, such as JT, Liouville, BF model, Chern-Simons gravity, and connected these universalities to the universal behavior of their partition functions, and also the universal behavior of Wheeler-DeWitt wave functions in these theories. We varied various parameters to track the changes in the spectrum, eigenfunctions, and partition functions and confirm the universal behaviors.

Next, we examined entanglement entropy in various wormhole solutions and using RG flow and the c-theorem, we extract further universalities in the structure of quantum gravity models. We also discussed complexity as a probe of RG flows in these wormhole geometries and showed that it can act as a tool to study these universal structures. Finally, we discussed the connections between topological defects and wormholes to further examine and confirm these RG flows and universalities.

Many questions remain to be answered, as we discussed throughout the work. Other models of quantum gravity and additional quantum information measures could specifically be used to extract and examine universalities. For instance, the kink solution of the $2d$ dilaton gravity model of \cite{Zhong:2023pel} could also be employed. The partition function of this model could be compared with the case of JT in order to generalize the universalities we observed here.

In \cite{Chen:2023mbc}, the contribution of the rotating black hole partition function, where ensemble equivalence is violated,  was discussed, as the behavior of thermal AdS and the rotating black hole are different. It would be interesting to check the change in the universalities we discussed here by considering this new rotating black hole saddle as well and to see how much of the would be preserved and how they would actually change.

Another important point is the presence of supersymmetry which, as explained in \cite{Turiaci:2023wrh}, while only fermionic modes are added, would completely change the spectrum of the theory and make it qualitatively different, as a gap is generated in supersymmetric theories due to the protection of ground state degeneracy by supersymmetry. Therefore, one would expect that supersymmetry only causes a gap shift in the behavior of the observables and universalities we tracked, which could be discussed further in future works.

In \cite{Jensen:2015swa}, it was shown that the central charge $b$, which after multiplying the Euler density, i.e., $ b \chi$ can characterize the CFT in higher dimensions, decreases monotonically or remain constant from UV to IR. Finding the connections between this parameter and associated wormhole structures would lead to interesting universal structures.  In particular, we can see that low-dimensional gravity universally loses accessible edge degrees of freedom along the RG flows.

 In \cite{Jensen:2015swa}, it was also found that for the Dirichlet boundary condition $b= - \frac{1}{16}$ and for the Neumann case $b = \frac{1}{16}$, where the RG flow is from Neumann boundary conditions with $b_{\text{UV}}=1/16$ to Dirichlet boundary conditions with $b_{IR}= - 1/16$. One interesting result there is the suggestion - as a general conjecture- that in a CFT, any observable can be reconstructed from an appropriate sum over ``conformal defects", similar to JT and other chaotic, low-dimensional gravity solutions. Thus, this defect central charge $b$ obeys a defect c-theorem.

 This result can also be rewritten in terms of the wormhole (or replica wormhole) formulation. Thus, the wormhole neck behaves as an emergent codimension-2 defect with a universal logarithmic term fixed by anomaly coefficients. These defect-Weyl-anomalies depend crucially on the extrinsic curvature invariants. Therefore, one could say that Liouville punctures, JT trumpets, matrix-model branes, and replica defects are all manifestations of one universal codimension-2 gravitational object, and wormholes may universally be understood as anomaly inflow channels between disconnected sectors, which could be studied further in future works.

\section*{Acknowledgments}

I would like to thank Imtak Jeon,  Alfredo Gonzalez Lezcano, and Augniva Ray for useful discussions.

 \medskip

\bibliography{LowdGrav.bib}
\bibliographystyle{JHEP}
\end{document}